%% file: outlier_HT_r7_nocolor.tex
\documentclass[onecolumn,10pt]{IEEEtran}
\input{preamble}

\usepackage{subfigure,epstopdf,bbm} 
\usepackage[utf8]{inputenc} 
\usepackage[T1]{fontenc}    
\usepackage{url,bbm}            
\usepackage{booktabs}       
\usepackage{amsfonts}       
\usepackage{nicefrac}       
\usepackage{microtype}      
\usepackage{cite}
\usepackage{subfigure,transparent,color,graphicx} 

\newcommand{\bbo}{\mathbbm{1}}
\allowdisplaybreaks[2]
\begin{document}
\title{Second-Order Asymptotically Optimal Outlier Hypothesis Testing}
\author{Lin Zhou, Yun Wei and Alfred Hero

\thanks{Lin Zhou is with the School of Cyber Science and Technology, Beihang University, Beijing, China, 100083 (Email: lzhou@buaa.edu.cn).} 
\thanks{Yun Wei is with the Department of Statistical Science, Duke University, Durham, NC 27708 (Email: yun.wei@duke.edu).}
\thanks{Alfred Hero is with the Department of Electrical Engineering and Computer Science, University of Michigan, Ann Arbor, MI, USA, 48109 (Email: hero@eecs.umich.edu).}

\thanks{This work was supported by the National Key Research and Development Program of China under Grant 2020YFB1804800, by a start up grant at the Beihang university, by ARO grants W911NF-15-1-0479 and W911NF-19-1-0269, and by DOE grant DE-NA0003921.}

\thanks{Copyright (c) 2017 IEEE. Personal use of this material is permitted.  However, permission to use this material for any other purposes must be obtained from the IEEE by sending a request to pubs-permissions@ieee.org.}

}

\maketitle

\begin{abstract}
We revisit the outlier hypothesis testing framework of Li \emph{et al.} (TIT 2014) and derive fundamental limits for the optimal test under the generalized Neyman-Pearson criterion. In outlier hypothesis testing, one is given multiple observed sequences, where most sequences are generated i.i.d. from a nominal distribution. The task is to discern the set of outlying sequences that are generated from anomalous distributions. The nominal and anomalous distributions are \emph{unknown}. We study the tradeoff among the probabilities of misclassification error, false alarm and false reject for tests that satisfy weak conditions on the rate of decrease of these error probabilities as a function of sequence length. Specifically, we propose a threshold-based test that ensures exponential decay of misclassification error and false alarm probabilities. We study two constraints on the false reject probability, with one constraint being that it is a non-vanishing constant and the other being that it has an exponential decay rate. For both cases, we characterize bounds on the false reject probability, as a function of the threshold, for each pair of nominal and anomalous distributions and demonstrate the optimality of our test under the generalized Neyman-Pearson criterion. We first consider the case of at most one outlying sequence and then generalize our results to the case of multiple outlying sequences where the number of outlying sequences is unknown and each outlying sequence can follow a different anomalous distribution.
\end{abstract}


\section{Introduction}

In the outlier hypothesis testing (OHT) problem, one observes a number $M$ of sequences. The majority of the $M$ sequences are i.i.d. samples from a nominal distribution and the rest of the sequences are i.i.d. samples from anomalous distributions different from the nominal distribution. The task in the  OHT problem is to design a test to discern the set of outlying sequences with high probability when both nominal and anomalous distributions are \emph{unknown}. Motivated by practical applications in anomaly detection~\cite{chandola2009anomaly}, we revisit the OHT problem studied in~\cite{li2014} when the outlying sequence might not be present and derive the performance tradeoff among the probabilities of misclassification error, false alarm and false reject for threshold-based tests. Furthermore, we show that such tests are optimal under the generalized Neyman-Pearson criterion for both a second-order asymptotic regime and a large deviations regime. Our second-order asymptotic result provides an approximation to the finite sample performance of the tests. Throughout the paper, we consider the case where the sequences have a finite alphabet.

We first consider the case when there is at most one outlying sequence. Under this setting, the null hypothesis is that there is no outlying sequence while a non-null hypothesis specifies the index of the outlying sequence. Li \emph{et al.}~\cite[Theorem 5]{li2014} showed that the error probability under each non-null hypothesis decays exponentially fast and that the error probability under the null hypothesis vanishes as the length of observed sequenced tends to infinity for the threshold based generalized likelihood ratio test~\cite[Eq. (25)]{li2014}. Furthermore, the authors of \cite{li2014} showed the optimality of their test when the number of observed sequences $M$ tends to infinity. A natural question arises: whether or not it is possible to claim optimality for a test when the number of observed sequences $M$ is \emph{finite} and when the length of the observed sequences is \emph{non-asymptotic.} Our first contribution sheds lights on the positive answer for this question. To do so, we decompose the error probability under the non-null hypothesis into the misclassification error probability and the false reject probability, where the false reject event corresponds to falsely claiming that no outlying sequence exists and the misclassification error event corresponds to falsely claiming that a nominal sequence is an outlier. The error probability under the null hypothesis is denoted the probability of false alarm, which is the probability of falsely claiming that an observed sequence is an outlying sequence when no outlying sequence is present. We show that a test, inspired by sequence classification with empirical statistics~\cite{gutman1989asymptotically,zhou2018binary}, is optimal under the generalized Neyman-Pearson criterion, from a second-order or a first-order asymptotic perspective.

We then generalize our results to the case where the number of outlying sequences is \emph{unknown} and each outlying sequence can be generated from a potentially different anomalous distribution. When the number of outlying sequences is known, Li \emph{et al.}~\cite[Theorem 10]{li2014} derived an achievability decay rate of the error probabilities under each hypothesis and showed asymptotic optimality of their result when the number of the sequences $M$ tends to infinity, when the lengths of sequences $n$ tend to infinity and when all the outlying sequences are generated from the same anomalous distribution. Furthermore, when the number of outlying sequences is unknown and when each outlying sequence is generated from the same anomalous distribution, Li \emph{et al.}~\cite[Theorem 10]{li2014} showed that when the null hypothesis is not taken into account, a generalized likelihood ratio test is exponentially consistent. However, the authors of \cite{li2014} did not provide explicit equations of the exponent. One might wonder whether it is possible to characterize the performance of a test when the number of outlying sequences is unknown and when each outlying sequence can be generated from a different anomalous distribution. Our second contribution provides a positive answer to this question and also demonstrates the optimality of the test under the generalized Neyman-Pearson criterion.

\subsection{Main Contributions}
Our main contribution is an analysis of the tradeoff among probabilities of misclassification error, false reject and false alarm for threshold-based tests that are optimal under the generalized Neyman-Pearson criterion~\cite{gutman1989asymptotically}. For the case where there exists at most one outlying sequence, our results complement \cite[Theorem 5]{li2014}, extending their results to a new threshold-based test and providing a second-order asymptotic approximation to the performance of the  test with finite length sequences. We also relax the conditions for optimality of the test using a weaker condition inspired by statistical classification~\cite{gutman1989asymptotically}. Furthermore, asymptotically, our results in Theorem \ref{result:exp1out} complement \cite[Proposition 4]{li2014} by identifying a sufficient condition on the pair of nominal and anomalous distributions under which the test ensures exponential decay of all three kinds of error probabilities. Finally, for the second-order asymptotic results in Theorem \ref{result:1outlier}, the information theoretical quantity that characterizes the performance tradeoff is shown to be a generalized Jensen-Shannon divergence, which is significantly different from the constrained sum of KL divergences in \cite[Eq. (26)]{li2014} or the Bhattacharyya distance in \cite[Corollary 6]{li2014}. For the case admitting multiple outlying sequences, we analyze the performance of a threshold-based test ignorant of the number of outlying sequences, where each outlying sequence can be drawn from a different distribution. Our results close a gap in the theory developed in \cite[Section IV]{li2014}, providing explicit equations for the asymptotic performance of the outlier test.

For the case where there exists at most one outlying sequence, we propose a threshold-based test that ensures exponential decay of both misclassification error and false alarm probabilities, called the homogeneous error exponent, which simultaneously upper bounds the false reject probabilities as a function of the threshold for any pair of nominal and anomalous distributions. We first derive a second-order asymptotic result that provides an approximation to the performance of the test when the length $n$ of each observed sequence is finite. In particular, under any pair of nominal and anomalous distributions $(P_\rmN,P_\rmA)$, we show that if the threshold of the test is upper bounded by a certain function of $n$, the false reject probability is essentially upper bounded by a constant $\varepsilon\in(0,1)$. Our proposed test is optimal under the generalized Neyman-Pearson criterion: among all tests that can ensure exponential decay of misclassification error and false alarm probabilities at a given rate for all pairs of nominal and anomalous distributions, our test has the smallest non-vanishing false reject probability under any pair of nominal and anomalous distributions. This way, optimality is ensured for any finite number of observed sequences $M$  (see \cite{zhou2018binary} for a similar result in the context of statistical classification).

In anomaly detection, it may be necessary to maintain a vanishingly small false reject probability when the length of each observed sequence becomes unbounded. To resolve this problem, asymptotically when the lengths of the observed sequences tend to infinity, we derive the exponential decay rate of the false reject probability as a function of the threshold in the test. We show, in particular, that the homogeneous error exponent is the threshold of the test for any pair of nominal and anomalous distributions. This way, we establish that, as long as the nominal and anomalous distribution is separated in a certain distance measure, the test is exponentially consistent, i.e., all three kinds of error probabilities decay exponentially fast with respect to the sequence length $n$. Conversely, we show that among all tests that can enure the same speed of exponential decay of misclassification error and false alarm probabilities for all pairs of nominal and anomalous distributions, our proposed test guarantees the largest exponential decay rate for the false reject probability regardless of the pair of nominal and anomalous distributions.

\subsection{Related Works}
The most closely related work to ours is that of \cite{li2014}, where the authors formulated the outlier hypothesis testing problem, and derived optimal results under constraints on the number of observed sequences, the length of observed sequences and the number of anomalous distributions. Other related work on outlier hypothesis testing is worth mentioning. A low complexity test for outlier hypothesis testing was proposed and analyzed in~\cite{bu2019linear}. A distribution free test based on maximum mean discrepancy was proposed in \cite{Zou2017TSP} and shown to be exponentially consistent when the number of outlying sequences is known, as long as a certain condition holds on the number of observed sequences and the length of each sequence. Readers may also refer to \cite{tajer2014outlying} for a comprehensive survey of the commonly made assumptions on distributions, definitions of outliers, types of tests and applications. Furthermore, the results of \cite{li2014} were generalized to a sequential setting in~\cite{li2017universal} where each sequence is observed symbol by symbol until the test is confident enough to make a decision. In \cite{Niti2015ISIT}, the authors studied the quickest outlier detection problem where outlying sequences follow an anomalous distribution after a certain unknown change time and tests were proposed to identify the outliers. Finally, in \cite{cohen2015active} for the problem of detecting an outlier from $M$ sequence streams, the authors studied a special case of the sequential outlier detection problem where at each time only a subset of all sequence symbols are observed.

Since our proof technique is inspired by asymptotic statistical classification theory, we also mention a few works in this domain. In \cite{gutman1989asymptotically}, Gutman studied a binary sequence classification problem and showed that a certain test using empirical distributions is asymptotically optimal with exponentially decreasing misclassification error probabilities. The result in \cite{gutman1989asymptotically} was generalized to classification of multiple sequences in~\cite{unnikrishnan2015asymptotically} and to distributed detection in~\cite{he2019distributed}. Finally, a finite sample analysis for the setting of~\cite{gutman1989asymptotically} was provided in~\cite{zhou2018binary}.

\subsection{Organization for the Rest of the Paper}
The rest of the paper is organized as follows. In Section \ref{sec:pf}, we set up the notation, formulate the outlier hypothesis testing problem with at most one outlying sequence, propose fundamental limits and  present our main results. In Section \ref{sec:Tout}, we generalize our results to the case of multiple outlying sequences where the number of outlying sequences is unknown and each outlying sequence is generated from a potentially different anomalous distribution. Finally, we conclude the paper and discuss future research directions in Section \ref{sec:conclude}. The proofs of all theorems are deferred to appendices.

\section{Case of At Most One Outlying Sequence}
\label{sec:pf}
\subsection*{Notation}

Random variables and their realizations are in upper (e.g.,  $X$) and lower case (e.g.,  $x$) respectively. All sets are denoted in calligraphic font (e.g.,  $\mathcal{X}$). We use superscripts to denote the vectors, like $X^n:=(X_1,\ldots,X_n)$. All logarithms are base $e$. The set of all probability distributions on a finite set $\calX$ is denoted as $\calP(\calX)$. Notation concerning the method of types follows~\cite{Tanbook}. Given a vector $x^n = (x_1,x_2,\ldots,x_n) \in\calX^n$, the {\em type} or {\em empirical distribution} is denoted as $\hat{T}_{x^n}(a)=\frac{1}{n}\sum_{i=1}^n \mathbbm{1}\{x_i=a\},a\in\calX$. The set of types formed from length-$n$ sequences with alphabet $\calX$ is denoted as $\calP_{n}(\calX)$. Given $P\in\calP_{n}(\calX)$, the set of all sequences of length $n$ with type $P$, the {\em type class}, is denoted as $\calT^n_P$. We use $\bbR$, $\bbR_+$, and $\bbN$ to denote the set of real numbers, non-negative real numbers, and  natural numbers respectively. Given any number $a\in\bbN$, we use $[a]$ to denote the collection of natural numbers between $1$ and $a$. 

\subsection{Problem Formulation}
We start by assuming that there is at most one outlying sequence. Consider a set of $M$ observed sequences $\bX^n:=\{X_1^n,\ldots,X_M^n\}$ and a pair of nominal distribution $P_\rmN$ and anomalous distribution $P_\rmA$ defined on the finite alphabet $\calX$. All sequences, with at most one exception, are generated i.i.d. from $P_\rmN$. The goal of outlier hypothesis testing is to discern the outlying sequence that is generated i.i.d. from the anomalous distribution $P_\rmA\in\calP(\calX)$ if an outlying sequence is present. Throughout this paper, we assume that both the nominal distribution $P_\rmN$ and the anomalous distribution $P_\rmA$ are \emph{unknown}. Furthermore, to avoid degenerate cases, similarly to \cite{li2014}, we consider only distributions $(P_\rmN,P_\rmA)$ with identical supports.

Under this setting, the objective of detecting a potential outlying sequence is equivalent to making a correct decision in the $(M+1)$-ary hypothesis testing problem with the following hypotheses:
\begin{itemize}
\item $\rmH_i,~i\in[M]$: the $i$-th sequence $X_i^n$ is the outlying sequence, i.e., $X_i^n\sim P_\rmA$ and $X_j^n\sim P_\rmN$ for all $j\in\calM_i$;
\item $\rmH_\rmr$: there is no outlying sequence, i.e., $X_j^n\sim P_\rmN$ for all $j\in[M]$,
\end{itemize}
where $\calM_i$ is defined the as the set of integers in $[M]$ excluding $i$ and $\rmH_\rmr$ denotes the null hypothesis.

The main task in the above OHT problem is to design a decision rule (test) $\phi_n:\calX^{Mn}\to\{\rmH_1,\ldots,\rmH_M,\rmH_\rmr\}$ having good performance in a sense specified below. Any test $\phi_n$ partitions the sample space $\calX^{Mn}$ into $M+1$ disjoint regions: $\{\calA_i(\phi_n)\}_{i\in[M]}$ where $X^{Mn}\in\calA_i(\phi_n)$ favors the non-null hypothesis $\rmH_i$ and a reject region $\calA_\rmr(\phi_n)=(\cup_{i\in[M]}\calA_i(\phi_n))^\rmc$ where $X^{Mn}\in\calA_i(\phi_n)$ favors the null hypothesis $\rmH_\rmr$.

Given any test $\phi_n$ and any pair of nominal and anomalous distributions $(P_\rmN,P_\rmA)\in\calP(\calX)^2$, the performance of the test $\phi_n$ is evaluated by the following misclassification error, false reject and false alarm probabilities:
\begin{align}
\beta_i(\phi_n|P_\rmN,P_\rmA)
&:=\bbP_i\{\phi_n(\bX^n)\notin\{\rmH_i,\rmH_\rmr\}\}\label{def:error},~i\in[M],\\
\zeta_i(\phi_n|P_\rmN,P_\rmA)
&:=\bbP_i\{\phi_n(\bX^n)=\rmH_\rmr\}\label{def:reject},~i\in[M],\\
\rmP_{\mathrm{fa}}(\phi_n|P_\rmN,P_\rmA)
&:=\bbP_\rmr\{\phi_n(\bX^n)\neq \rmH_\rmr\}
\label{def:falarm},
\end{align}
where for each $i\in[M]$, we define $\bbP_i\{\cdot\}:=\Pr\{\cdot|\rmH_i\}$ where $X_i^n$ is distributed i.i.d. according to $P_\rmA$ and for $X_j^n$ is distributed according to $P_\rmN$ for each $j\in\calM_i$ and we define $\bbP_\rmr\{\cdot\}:=\Pr\{\cdot|\rmH_\rmr\}$ where all sequences are generated i.i.d. from $P_\rmN$ for all $i\in[M]$. Consistent with the literature on hypothesis testing~(e.g., \cite{lehmann2006testing}), we define $\beta_i(\phi_n|P_\rmN,P_\rmA)$ and $\zeta_i(\phi_n|P_\rmN,P_\rmA)$ as type-$i$ misclassification error and false reject probabilities, respectively, and we define $\rmP_{\mathrm{fa}}(\phi_n|P_\rmN,P_\rmA)$ as the false alarm probability. Our main results characterize the tradeoff among the probabilities of misclassification error in \eqref{def:error}, false rejection in \eqref{def:reject} and false alarm in \eqref{def:falarm} for different settings.

\subsection{A Threshold-Based Test}
To present our test, we need the following definition. Given a sequence of distributions $\bQ=(Q_1,\ldots,Q_M)\in\calP(\calX)^M$, for each $i\in[M]$, define the following linear combination of KL divergence terms between a single distribution and a mixture distribution
\begin{align}
\rmG_i(\bQ)
&:=\sum_{j\in\calM_i}D\left(Q_j\bigg\|\frac{\sum_{l\in\calM_i}Q_l}{M-1}\right)\label{def:gi},
\end{align}
where $\calM_i$ was defined the as the set of integers in $[M]$ excluding $i$. We remark that $\rmG_i(\bQ)$ can be understood as a homogeneity measure that checks the similarity of distributions $\bQ$ except $Q_i$. The measure $\rmG_i(\bQ)=0$ if and only if $Q_j=Q$ for all $j\in\calM_i$ where $Q\in\calP(\calX)$ is arbitrary. This measure  will be used to construct our optimal test.

Throughout the section, we use a threshold-based test that takes the observed sequences as inputs and it outputs a decision among the $(M+1)$ hypotheses. Given $M$ observed sequences $\bx^n=(x_1^n,\ldots,x_M^n)$ and any positive real number $\lambda\in\bbR_+$, the test operates as follows:
\begin{align}
\psi_n(\bx^n)
&:=\left\{
\begin{array}{cc}
\rmH_i&\mathrm{if}~\rmS_i(\bx^n)<\min_{j\in\calM_i}\rmS_j(\bx^n)\mathrm{~and~}\min_{j\in\calM_i}\rmS_j(\bx^n)>\lambda\\
\rmH_\rmr&\mathrm{otherwise},
\end{array}
\right.\label{test1outlier}
\end{align}
where $\rmS_i(\bx^n)$ is the scoring function
\begin{align}
\rmS_i(\bx^n)&:=\rmG_i(\hatT_{x_1^n},\ldots,\hatT_{x_M^n})\label{def:scoref},
\end{align}
and $\rmG_i(\cdot)$ is the function defined in \eqref{def:gi} that measures the sum of the KL divergence between the empirical distribution of each sequence $x_j^n$ with $j\in\calM_i$ relative to the average of the empirical distributions of all sequences $x_j^n$ where $j\in\calM_i$. Note that the threshold $\lambda$ may be a function of sequence length $n$, denoted $\lambda_n$ discussed below.

We first informally explain the test in \eqref{def:scoref} from an asymptotic point of view. Intuitively, if $x_i^n$ is the anomalous sequence that is generated from the unknown distribution $P_\rmA$, then as the length of each observed sequence $n$ increases, using the weak law of large numbers, we know that the empirical distribution $\hatT_{x_i^n}$ tends to $P_\rmA$ and the empirical distribution $\hatT_{x_j^n}$ for each $j\in\calM_i$ tends to the unknown nominal distribution $P_\rmN$. Thus, the scoring function $\rmS_i(\bx^n)$ tends to zero and the scoring function of $\rmS_j(\bx^n)$ for each $j\in\calM_i$ tends to $\mathrm{GD}_M(P_\rmN,P_\rmA)$ (cf. \eqref{def:GD}), which is strictly positive if $P_\rmN\neq P_\rmA$. Therefore, for any threshold $\lambda$ that is positive but less than $\mathrm{GD}_M(P_\rmN,P_\rmA)$, with high probability, it is possible to identify the outlying sequence if it exists. On the other hand, if there is no outlier, then with the same logic, for each $i\in[M]$, the scoring function $\rmS_i(\bx^n)$ tends to zero and naturally the null hypothesis is decided for any positive threshold $\lambda$. Therefore, the test in \eqref{test1outlier} is consistent asymptotically for any $P_\rmN\neq P_\rmA$ such that the threshold $\lambda<\mathrm{GD}_M(P_\rmN,P_\rmA)$.

We remark that the scoring function $\rmS_i(\bx^n)$ was also used in \cite[Eq. (25)]{li2014} to construct a test for the same problem. At first glance, the threshold-based test in \cite[Eq. (25)]{li2014} relies on the pairwise difference of log likelihoods of the joint empirical distributions under different hypotheses. However, a closer investigation reveals that the test in \cite[Eq. (25)]{li2014} is equivalent to the following test
\begin{align}
\psi_n^{\rm{Li}}(\bx^n)
&:=\left\{
\begin{array}{cc}
\rmH_i&\mathrm{if}~\rmS_i(\bx^n)<\min_{j\in\calM_i}S_j(\bx^n)\mathrm{~and~}\max_{j\neq k} (S_j(\bx^n)-S_k(\bx^n))>\alpha_n,\\
\rmH_\rmr&\mathrm{otherwise},
\end{array}
\right.
\end{align}
where $\alpha_n=\Theta(\frac{\log n}{n})$ is a length-$n$ dependent threshold. In \cite[Theorem 5]{li2014}, Li \emph{et. al} showed that their test $\psi_n^{\rm{Li}}$ ensures that the sum of false reject and misclassification error probabilities decay exponentially fast and that the false alarm probability vanishes as $n\to\infty$. Note that our test in \eqref{test1outlier} differs from the test \cite[Eq. (25)]{li2014} only in that 
we have a different condition to decide the null hypothesis. However, this subtle difference enables us to trade off the false reject probability and the homogeneous decay rate of the misclassification error and false alarm probabilities in Theorems \ref{result:1outlier} to \ref{result:exp1out}. It should be noted that, while the test $\psi_n^{\rm{Li}}$ is universal over $(P_\rmN,P_\rmA)$, this was only achievable since the false alarm probability was not controlled in \cite{li2014}. The false alarm control in our proposed test will necessarily depend on the knowledge of $P_\rmN$, through the threshold $\lambda$, and therefore our test in \eqref{test1outlier} is not universal.

\subsection{Preliminaries}
\label{sec:prelim}
To present our results that characterize the tradeoff among the probabilities of misclassification error, false alarm and false reject, several definitions are needed. Given any pair of distributions $(P_\rmN,P_\rmA)$, for any $x\in\calX$, define two information densities (log likelihood ratios):
\begin{align}
\imath_1(x|P_\rmN,P_\rmA)
&:=\log\frac{(M-1)P_\rmA(x)}{(M-2)P_\rmN(x)+P_\rmA(x)},\label{def:id1}\\
\imath_2(x|P_\rmN,P_\rmA)
&:=\log\frac{(M-1)P_\rmN(x)}{(M-2)P_\rmN(x)+P_\rmA(x)}\label{def:id2}.
\end{align}
The following linear combinations of the expectations and variances of these two information densities are critical in presenting our main results:
\begin{align}
\mathrm{GD}_M(P_\rmN,P_\rmA)
&:=\mathbb{E}_{P_\rmA}[\imath_1(X|P_\rmN,P_\rmA)]+(M-2)\mathbb{E}_{P_\rmN}[\imath_2(X|P_\rmN,P_\rmA)],\label{def:GD}\\
\rmV_M(P_\rmN,P_\rmA)
&:=\mathrm{Var}_{P_\rmA}[\imath_1(X|P_\rmN,P_\rmA)]+(M-2)\mathrm{Var}_{P_\rmN}[\imath_2(X|P_\rmN,P_\rmA)]\label{def:V}.
\end{align}
Furthermore, we need the following covariance function of the information densities
\begin{align}
\mathrm{Cov}_M(P_\rmN,P_\rmA)
\nn&:=-\big(\mathrm{GD}_M(P_\rmN,P_\rmA)\big)^2+\mathbb{E}_{P_\rmA}\Big[\big(\imath_1(X|P_\rmN,P_\rmA)\big)^2\Big]+2(M-2)\mathbb{E}_{P_\rmA}[\imath_1(X|P_\rmN,P_\rmA)]\mathbb{E}_{P_\rmN}[\imath_2(X|P_\rmN,P_\rmA)]\\
&\qquad+(M^2-5M+7)\big(\mathbb{E}_{P_\rmN}[\imath_2(X|P_\rmN,P_\rmA)]\big)^2+(M-3)\mathbb{E}_{P_\rmN}\Big[\big(\imath_2(X|P_\rmN,P_\rmA)\big)^2\Big]\label{def:cov}.
\end{align}
Then, the covariance matrix $\bV_M(P_\rmN,P_\rmA)=\{V_{i,j}(P_\rmN,P_\rmA)\}_{(i,j)\in[M-1]^2}$ is defined as
\begin{align}
V_{i,j}(P_\rmN,P_\rmA)
&=\left\{
\begin{array}{ll}
\rmV_M(P_\rmN,P_\rmA)&\mathrm{if~}i=j\\
\mathrm{Cov}_M(P_\rmN,P_\rmA)&\mathrm{otherwise}.
\end{array}
\right.
\label{defdispersionmatrix}
\end{align}
For any $k\in\bbN$, $\rmQ_k(x_1,\ldots,x_k;\bmu,\bSigma)$ is the multivariate generalization of the complementary Gaussian cdf defined as follows:
\begin{align}
\rmQ_k(x_1,\ldots,x_k;\bmu,\bSigma)
&:=\int^{\infty}_{x_1}\ldots\int^{\infty}_{x_k}\calN(\bx;\bmu;\bSigma)\rmd \bx\label{def:kQ},
\end{align}
where $\calN(\bx; \bmu;\bSigma)$ is the pdf of a $k$-variate Gaussian with mean $\bmu$ and covariance matrix $\bSigma$~\cite{tan2015asymptotic}. Furthermore, for any $k\in\bbN$, we use $\mathbf{1}_k$ to denote a row vector of length $k$ with all elements being one and we use $\mathbf{0}_k$ similarly. The complementary Gaussian cdf with covariance matrix $\bV_M(P_\rmN,P_\rmA)$ and the mean value $\mathrm{GD}_M(P_\rmN,P_\rmA)$ bounds the probability of false reject.

Finally, given any $\lambda\in\bbR_+$ and any pair of distributions $(P_\rmN,P_\rmA)$, for each $i\in[M]$, define the following quantity
\begin{align}
\mathrm{LD}_i(\lambda,P_\rmN,P_\rmA)
&:=\min_{(j,k)\in[M]^2:j\neq k}\min_{\substack{\bQ\in(\calP(\calX))^M:\\\rmG_j(\bQ)\leq\lambda,~\rmG_k(\bQ)\leq\lambda}}\Big(D(Q_i\|P_\rmA)+\sum_{l\in\calM_i}D(Q_l\|P_\rmN)\Big)\label{def:LDi}.
\end{align}
The above quantity is key to characterize the exponential decay rate of the false reject probability.

\subsection{Second-Order Asymptotic Approximation to the Non-Asymptotic Performance}
Our first set of results characterize the performance of the test in \eqref{test1outlier} in terms of probabilities of misclassification error, false alarm and false reject probabilities in the second-order asymptotic regime. We first demonstrate a non-asymptotic achievability result and then prove that such a result is optimal up to second-order under the generalized Neyman-Pearson criterion~\cite{gutman1989asymptotically,zhou2018binary}.

\subsubsection{Achievability}
\begin{theorem}
\label{result:1outlier}
For every pair of nominal and anomalous distributions $(P_\rmN,P_\rmA)\in\calP(\calX)^2$, given any positive real number $\lambda\in\bbR_+$, the test in \eqref{test1outlier} satisfies
\begin{align}
\max_{i\in[M]}\beta_i(\psi_n|P_\rmN,P_\rmA)&\leq \exp\Big(-n\lambda+|\calX|\log((M-1)n+1)\Big),\\
\rmP_{\mathrm{fa}}(\psi_n|P_\rmN,P_\rmA)&\leq M(M-1)\exp(-n\lambda+|\calX|\log((M-1)n+1)),\\
\max_{i\in[M]}\zeta_i(\psi_n|P_\rmN,P_\rmA)
&\leq 1-\rmQ_{M-1}\bigg(\sqrt{n}\Big(\lambda-\mathrm{GD}_M(P_\rmN,P_\rmA)+O\Big(\frac{\log n}{n}\Big)\Big)\times\mathbf{1}_{M-1};\mathbf{0}_{M-1};\bV_M(P_\rmN,P_\rmA)\bigg)+O\left(\frac{1}{\sqrt{n}}\right)\label{upp3}.
\end{align}
\end{theorem}
The proof of Theorem \ref{result:1outlier} is provided in Appendix \ref{proof:result:1outlier}. We make several remarks.

For any finite number of observed sequences $M$, as the length of each observed sequence $n$ increases, both the maximal classification error (cf. \eqref{def:error}) and the false alarm (cf. \eqref{def:falarm}) probabilities decay exponentially fast with a speed lower bounded by the threshold $\lambda$ in the test in \eqref{test1outlier}, i.e.,
\begin{align}
\liminf_{n\to\infty}\frac{1}{n}\min\left\{\min_{i\in[M]}\{-\log\beta_i(\psi_n|P_\rmN,P_\rmA)\},-\log\rmP_{\mathrm{fa}}(\psi_n|P_\rmN,P_\rmA)\right\}\geq \lambda\label{asymplower}.
\end{align}
Furthermore, asymptotically, the upper bound on the maximal false reject probability (cf. \eqref{def:reject}) converges to $\lim_{n\to\infty}\big(1-\rmQ_{M-1}(\sqrt{n}(\lambda-\mathrm{GD}_M(P_\rmN,P_\rmA))\times\mathbf{1}_{M-1};\mathbf{0}_{M-1};\bV_M(P_\rmN,P_\rmA))\big)$, which is a function of the threshold $\lambda$ and the pair of distributions $(P_\rmN,P_\rmA)$. To better understand the seemingly complicated upper bound on the false reject probability, for any $\varepsilon\in(0,1)$, we define 
\begin{align}
L_M^*(\varepsilon|P_\rmN,P_\rmA)
&:=\max\Big\{L\in\bbR:\rmQ_{M-1}(L\times\mathbf{1}_{M-1};\mathbf{0}_{M-1};\bV_M(P_\rmN,P_\rmA))\geq 1-\varepsilon\Big\}\label{def:L*PNPA},\\
\lambda^*(n,\varepsilon|P_\rmN,P_\rmA)&:=\mathrm{GD}_M(P_\rmN,P_\rmA)+\frac{L_M^*(\varepsilon|P_\rmN,P_\rmA)}{\sqrt{n}}\label{def:lambda^*}.
\end{align}

\begin{figure}[tb]
\centering
\includegraphics[width=.5\columnwidth]{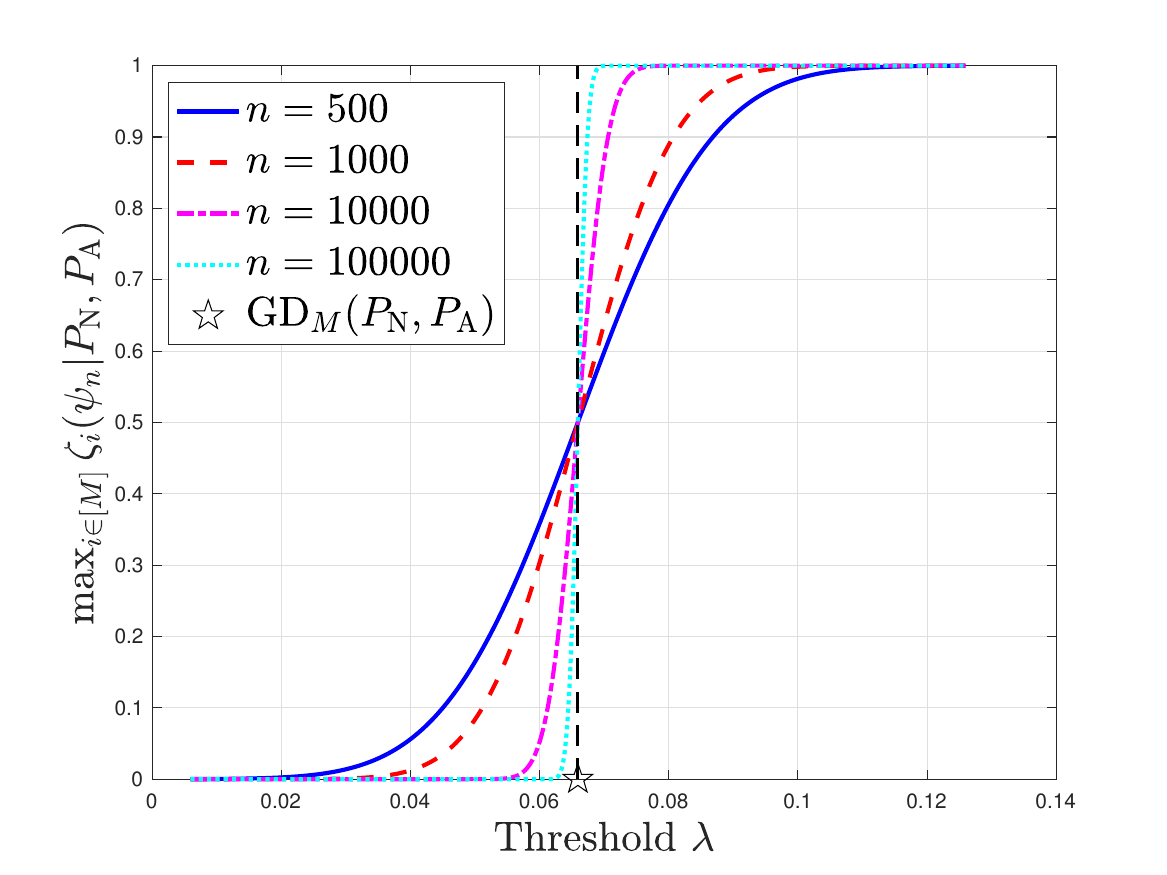}
\caption{
Illustration of phase transition for our test. Here we consider Bernoulli sources $P_\rmN=\mathrm{Bern}(0.2)$ and $P_\rmA=\mathrm{Bern}(0.4)$. We assume that there are $M=4$ observed sequences and one sequence is the outlier. We plot the maximal false reject probability, i.e., $\max_{i\in[n]}\zeta_i(\psi_n|P_\rmN,P_\rmA)$.}
\label{phasetransition}
\end{figure}

We then have the following corollary of Theorem \ref{result:1outlier}.
\begin{corollary}
\label{coro:1outlier}
For any pair of nominal and anomalous distributions $(P_\rmN,P_\rmA)$, if the threshold $\lambda$ satisfies $\lambda\leq \lambda^*(n,\varepsilon|P_\rmN,P_\rmA)$ for all $n\in\bbN$, then for any $\varepsilon\in(0,1)$, the maximal false reject probability under $(P_\rmN,P_\rmA)$ is asymptotically upper bounded by $\varepsilon$, i.e., $\limsup_{n\to\infty}\max_{i\in[M]}\zeta_i(\psi_n|P_\rmN,P_\rmA)\leq \varepsilon$. In particular, if the threshold $\lambda$ further satisfies that $\lambda<\mathrm{GD}_M(P_\rmN,P_\rmA)$ for all $n\in\bbN$, then the false reject probability vanishes. 
\end{corollary}
The result in Corollary \ref{coro:1outlier} implies a phase transition phenomenon for our test. In particular, if the threshold $\lambda$ is strictly greater than $\mathrm{GD}_M(P_N,P_A)$, then asymptotically the false reject probabilities tend to one. On the other hand, if $\lambda<\mathrm{GD}_M(P_N,P_A)$, then asymptotically the false reject probabilities vanish. See Figure \ref{phasetransition} for a numerical illustration. As we shall show later (in Theorem \ref{result:exp1out}), actually, if $\lambda<\mathrm{GD}_M(P_\rmN,P_\rmA)$, the false reject probability converges to zero \emph{exponentially} fast with a speed lower bounded by a explicit function of the threshold $\lambda$. 

Note that $\lambda^*(n,\varepsilon|P_\rmN,P_\rmA)$ is a critical bound for the threshold in the test, which trades off a lower bound $\lambda$ on the exponential decay rates of misclassification error and false alarm probabilities and a non-vanishing upper bound $\varepsilon\in(0,1)$ for the maximal false reject probability. Such a result is known as a second-order asymptotic result since it provides a formula for the second dominant term $\frac{L_M^*(\varepsilon|P_\rmN,P_\rmA)}{\sqrt{n}}$ beyond the leading constant term $\mathrm{GD}_M(P_\rmN,P_\rmA)$ asymptotically as $n\to\infty$. Furthermore, as shown in non-asymptotic analysis for channel coding~\cite{polyanskiy2010finite}, second-order asymptotic results often provide good approximation to the performance for finite length $n$. We provide a numerical example to illustrate the validity of this claim in Section \ref{sec:numerical}. 

\begin{figure}[tb]
\centering
\includegraphics[width=.5\columnwidth]{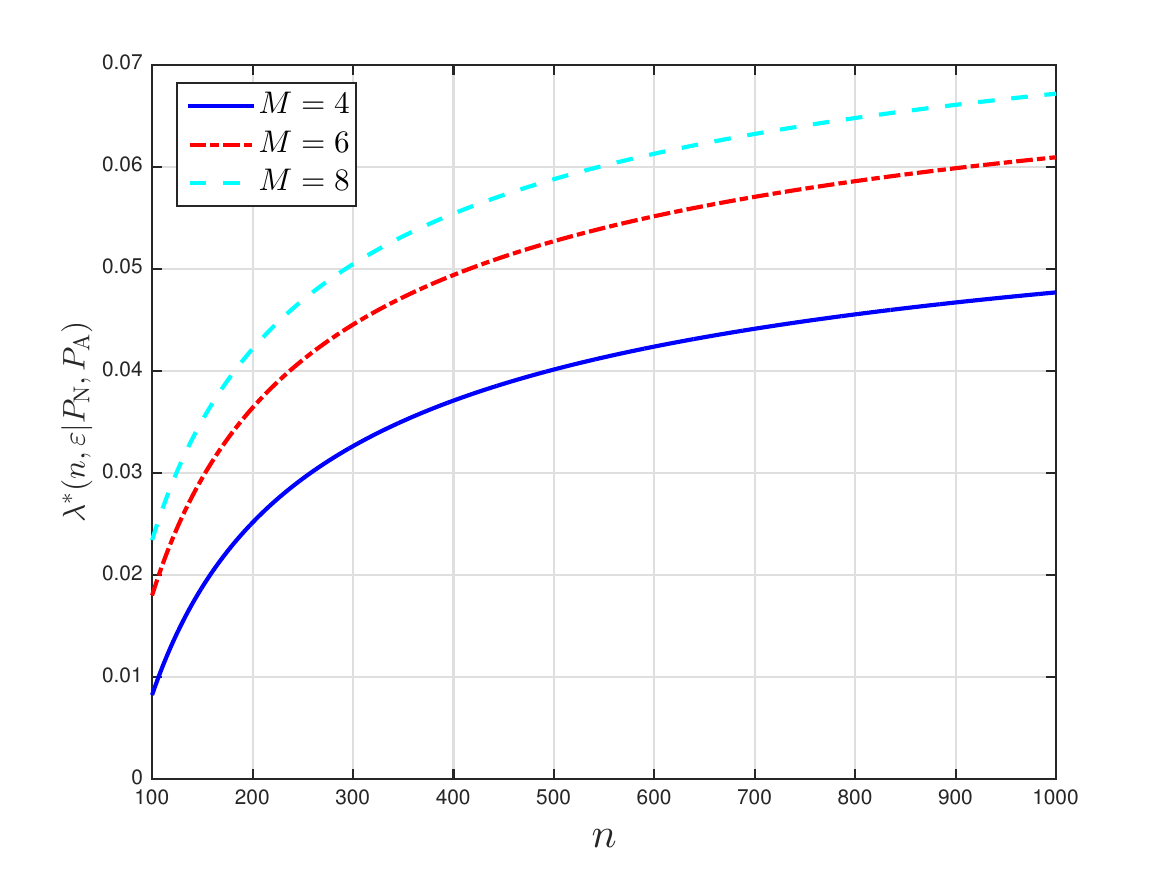}
\caption{
Illustration of the effect of the number of sequences $M$ in the performance for outlying sequence detection with reject option. Here we consider Bernoulli sources $P_\rmN=\mathrm{Bern}(0.2)$ and $P_\rmA=\mathrm{Bern}(0.4)$. We assume that there are $M$ observed sequences and that only one sequence is an outlier. We plot $\lambda^*(n,\varepsilon|P_\rmN,P_\rmA)$ without the $O(\log n/n)$ term.}
\label{maffect}
\end{figure}

Theorem \ref{result:1outlier} also captures the influence of the number of sequences $M$ on the performance of the test~\eqref{test1outlier}. To study the asymptotic case of $M\to\infty$, we need to make an assumption on the order of $M$ and $n$. In fact, as long as $\limsup_{n\to\infty}\frac{M\log n}{n}\to 0$, the asymptotic lower bounds hold. Intuitively, when one has a larger number of sequences, it should be easier to learn the nominal distribution and thus achieve better performance. This should imply that as $M$ increases, the upper bound $\lambda^*(n,\varepsilon|P_\rmN,P_\rmA)$ on the homogeneous error exponent $\lambda$ in \eqref{def:lambda^*} increases as well. To verify this intuition, the second-order result in \eqref{def:lambda^*} for Bernoulli distributions with different values of $M$ is plotted in Figure \ref{maffect}.
The influence of $M$ on the performance of the test in \eqref{test1outlier} is dominated by $\mathrm{GD}_M(P_\rmN,P_\rmA)$. In fact, 
\begin{align}
\frac{\partial \mathrm{GD}_M(P_\rmN,P_\rmA)}{\partial M}
&=D\left(P_\rmN\bigg\|\frac{(M-2)P_\rmN+P_\rmA}{M-1}\right)>0.
\end{align}
Thus, as the number of sequences $M$ increases, the performance of the test in \eqref{test1outlier} improves. In the extreme case, as $M\to\infty$, we have
\begin{align}
\lim_{M\to\infty}\mathrm{GD}_M(P_\rmN,P_\rmA)=D(P_\rmA\|P_\rmN).
\end{align}
This implies that the maximum asymptotic decay rate of the misclassification error and false alarm probabilities of the test under any pair of nominal and anomalous distributions $(P_\rmN,P_\rmA)$ in \eqref{test1outlier} is $D(P_\rmA\|P_\rmN)$ as the number of sequences $M$ tends to infinity, assuming that the false reject probability does not tend to \emph{one}.

Finally, we remark that Theorem \ref{result:1outlier} is relevant to $M$-ary hypothesis testing using empirical statistics~\cite{gutman1989asymptotically,unnikrishnan2015asymptotically,zhou2018binary}, also known as $M$-ary statistical classification. In $M$-ary statistical classification, one is given $M$ training sequences and one testing sequence. The task there is to identify the true distribution of the testing sequence among the empirical distributions of the training sequences. In contrast, in the outlier hypothesis testing problem addressed in Theorem \ref{result:1outlier}, we are given $M$ sequences and our task is to identify the potential outlying sequence if it exists. Although the two problems are different in formulation, the proof techniques are similar. In fact, our proof technique for Theorem \ref{result:1outlier} can be used to strengthen \cite[Theorem 4.1]{zhou2018binary} by removing the condition in \cite[Section 4.2]{zhou2018binary} on the uniqueness of the minimizing distribution for the scoring function in \cite[Eq. (4.4)]{zhou2018binary}.

\subsubsection{Converse}
With the above achievability result on the performance of the test in \eqref{test1outlier}, it remains to show that the test is in fact optimal in a certain sense. Since nominal and anomalous distributions are unknown, in order to derive a converse result, the classical Neyman-Pearson criterion, which requires knowledge of generating distributions, is not applicable. Furthermore, as proved in \cite{li2014}, for our problem, it is impossible to ensure that all three kinds of error probabilities decay exponentially for all pairs of nominal and anomalous distributions. As a compromise, we adopt the generalized Neyman-Pearson criterion of Gutman~\cite{gutman1989asymptotically} to derive a lower bound on the false reject probability. The generalized Neyman-Pearson criterion is that both misclassification error and false alarm probabilities decay exponentially fast with homogeneous speed for \emph{all} pairs of nominal and anomalous distributions. We give a lower bound on the false reject probability for any particular pair of distributions $(P_\rmN,P_\rmA)$ in the following theorem.
\begin{theorem}
\label{converse:1outlier}
Given any positive real number $\lambda\in\bbR_+$, let the test $\phi_n$ satisfy
\begin{align}
\max\Big\{\max_{i\in[M]}\beta_i(\phi_n|\tilP_\rmN,\tilP_\rmA),P_{\mathrm{fa}}(\phi_n|\tilP_\rmN,\tilP_\rmA)\Big\}\leq \exp(-n\lambda),~\forall~(\tilP_\rmN,\tilP_\rmA)\in\calP(\calX)^2.
\end{align}
Then for any pair of nominal and anomalous distributions $(P_\rmN,P_\rmA)\in\calP(\calX)^2$, the minimal false reject probability satisfies
\begin{align}
\min_{i\in[M]}\zeta_i(\phi_n|P_\rmN,P_\rmA)
&\geq 1-\rmQ_{M-1}\bigg(\sqrt{n}\Big(\lambda-\mathrm{GD}_M(P_\rmN,P_\rmA)+O\Big(\frac{\log n}{n}\Big)\Big)\times\mathbf{1}_{M-1};\mathbf{0}_{M-1};\bV_M(P_\rmN,P_\rmA)\bigg)+O\left(\frac{1}{\sqrt{n}}\right).
\end{align}
\end{theorem}
The proof of Theorem \ref{converse:1outlier} is provided in Appendix \ref{proof:converse}.

The result in Theorem \ref{converse:1outlier} holds for any number of observed sequences $M$ and when the length $n$ of each observed sequence $n$ is such that $O(\frac{\log n}{n})$ and $O(\frac{1}{\sqrt{n}})$ can be neglected. Furthermore, Theorem \ref{converse:1outlier} implies that the test in \eqref{test1outlier} is optimal under the generalized Neyman-Pearson criterion. Specifically, among all tests that ensure exponential decay of the maximal misclassification error and false alarm probabilities at a speed no less than $\lambda$, the test in \eqref{test1outlier} achieves the minimal false reject probability in a second-order asymptotic sense such that $\liminf_{n\to\infty}\rmQ_{M-1}\big(\sqrt{n}(\lambda-\mathrm{GD}_M(P_\rmN,P_\rmA)+O\Big(\frac{\log n}{n}))\times\mathbf{1}_{M-1};\mathbf{0}_{M-1};\bV_M(P_\rmN,P_\rmA)\big)>0$.

\subsubsection{A Numerical Example}
\label{sec:numerical}
We present an example to illustrate Theorem \ref{result:1outlier} and Corollary \ref{coro:1outlier}. Consider the binary alphabet $\calX=\{0,1\}$ and $M=4$. Assume that there is exactly one outlying sequence and let $\mathrm{Bern}(p)$ denote a Bernoulli distribution with parameter $p\in(0,1)$. For any $(p,q)\in(0,1)^2$ such that $p\neq q$, we set the nominal distribution $P_\rmN$ as $\mathrm{Bern}(p)$ and the anomalous distribution $P_\rmA$ as $\mathrm{Bern}(q)$. We make the above nominal and anomalous distribution assumptions in order to demonstrate tightness of the inequality \eqref{def:lambda^*} in the theorem. For the above example, the information densities (cf. \eqref{def:id1} and \eqref{def:id2}) satisfy
\begin{align}
\imath_1(x|P_\rmN,P_\rmx)
&:=\bbo(x=0)\log\frac{(M-1)(1-q)}{(M-2)(1-p)+1-q}+\bbo(x=1)\log\frac{(M-1)q}{(M-2)p+q},\\
\imath_2(x|P_\rmN,P_\rmx)
&:=\bbo(x=0)\log\frac{(M-1)(1-p)}{(M-2)(1-p)+1-q}+\bbo(x=1)\log\frac{(M-1)p}{(M-2)p+q}.
\end{align}
Furthermore, 
\begin{align}
\mathrm{GD}_M(P_\rmN,P_\rmA)
&=D_b\bigg(q\Big\|\frac{(M-2)p+q}{M-1}\bigg)+(M-2)D_b\bigg(p\Big\|\frac{(M-2)p+q}{M-1}\bigg),
\end{align}
where $D_b(p\|q)=p\log\frac{p}{q}+(1-p)\log\frac{1-p}{1-q}$ is the binary KL divergence function.
The variance $\rmV_M(P_\rmN,P_\rmA)$ is given by
\begin{align}
\rmV_M(P_\rmN,P_\rmA)
\nn&=\mathbb{E}_{P_\rmA}[(\imath_1(X|P_\rmN,P_\rmx))^2]+(M-2)\mathbb{E}_{P_\rmN}[(\imath_2(X|P_\rmN,P_\rmA))^2]\\*
&\qquad-\bigg(D_b\bigg(q\Big\|\frac{(M-2)p+q}{M-1}\bigg)\bigg)^2-(M-2)\bigg(D_b\bigg(p\Big\|\frac{(M-2)p+q}{M-1}\bigg)\bigg)^2.
\end{align}
Similarly, we can also calculate $\mathrm{Cov}_M(P_\rmN,P_\rmA)$ (cf. \eqref{def:cov}) and thus the covariance matrix $\bV_M(P_\rmN,P_\rmA)$.

For the case of $p=0.2$, $q=0.4$ and $M=4$, we have
\begin{align}
\bV_M(P_\rmN,P_\rmA)
&=\left[
\begin{array}{cccc}
0.1331&0.1106&0.1106\\
0.1106&0.1331&0.1106\\
0.1106&0.1106&0.1331
\end{array}
\right],
\end{align}
and other cases can be calculated similarly.

\begin{figure}[tb]
\begin{tabular}{cc}
\includegraphics[width=.5\columnwidth]{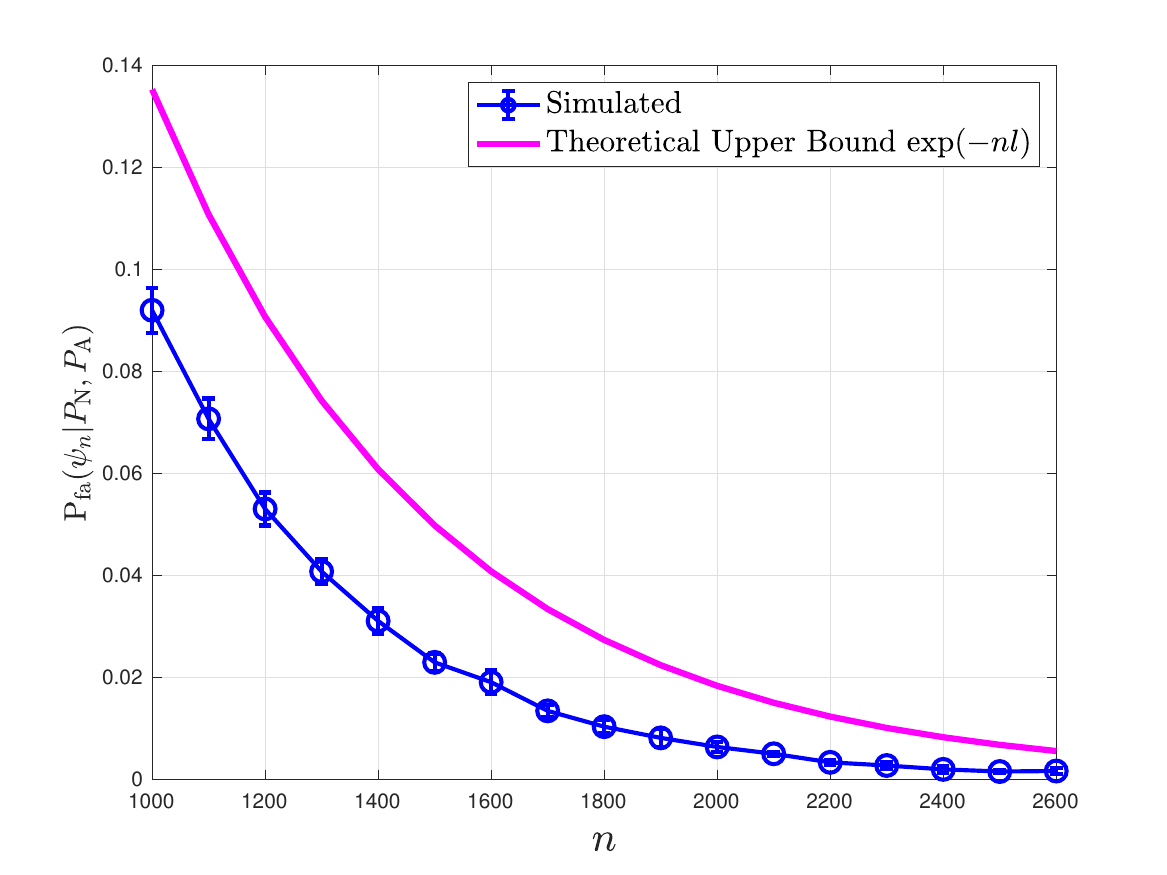}&\includegraphics[width=.5\columnwidth]{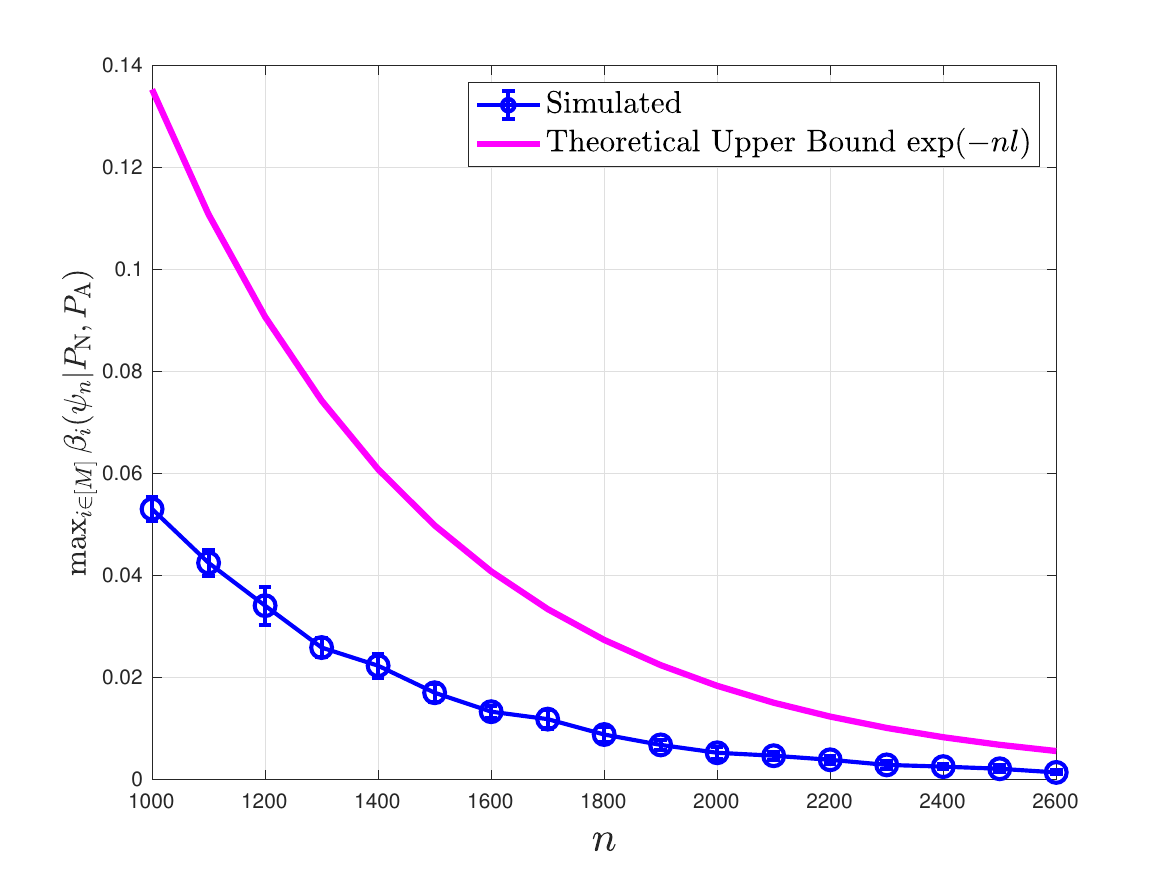}\\
{(a) False alarm probability} & {(b) Misclassification error probability}
\end{tabular}
\caption{Simulated false alarm and misclassification error probabilities for $M=4$. The false alarm probability is simulated for $P_\rmN=\mathrm{Bern}(0.25)$ and the misclassification error probability is simulated for $P_\rmN=\mathrm{Bern}(0.29)$ and $P_\rmA=\mathrm{Bern}(0.3)$. The error bar denotes one standard deviation below and above the mean value. As observed, both false alarm and misclassification error probabilities are upper bounded by the dominant term $\exp(-n\lambda)$ in Theorem \ref{result:1outlier}.}
\label{falarm}
\end{figure}

Below we simulate the false alarm and misclassification error probabilities of our test in \eqref{test1outlier} with $\lambda=0.002$\footnote{Such a choice of $\lambda$ is selected to ensure that $\exp(-nl)$ can be numerically approximated without excessive simulation trials.}. The false alarm probability is simulated for $P_\rmN=\mathrm{Bern}(0.25)$ and the misclassification error probability is simulated for $P_\rmN=\mathrm{Bern}(0.29)$ and $P_\rmA=\mathrm{Bern}(0.3)$. The false alarm probability is determined by the nominal distribution $P_\rmN$. The simulation results are plotted and compared with the theoretical upper bounds in Figure \ref{falarm}.  For each sequence length $n$, we run the test $10^6$ times and plot the empirical false reject probability. From Figure \ref{falarm}, we observe that both false alarm and misclassification error probabilities are upper bounded by $\exp(-n\lambda)$, which is the dominant term in the upper bounds derived in Theorem \ref{result:1outlier}. Thus, the simulation results in Figure \ref{falarm} demonstrate that our lower bound on the homogeneous decay rate of misclassification error and false alarm probabilities is valid for this numerical example.

\begin{figure}[tb]
\centering
\includegraphics[width=.5\columnwidth]{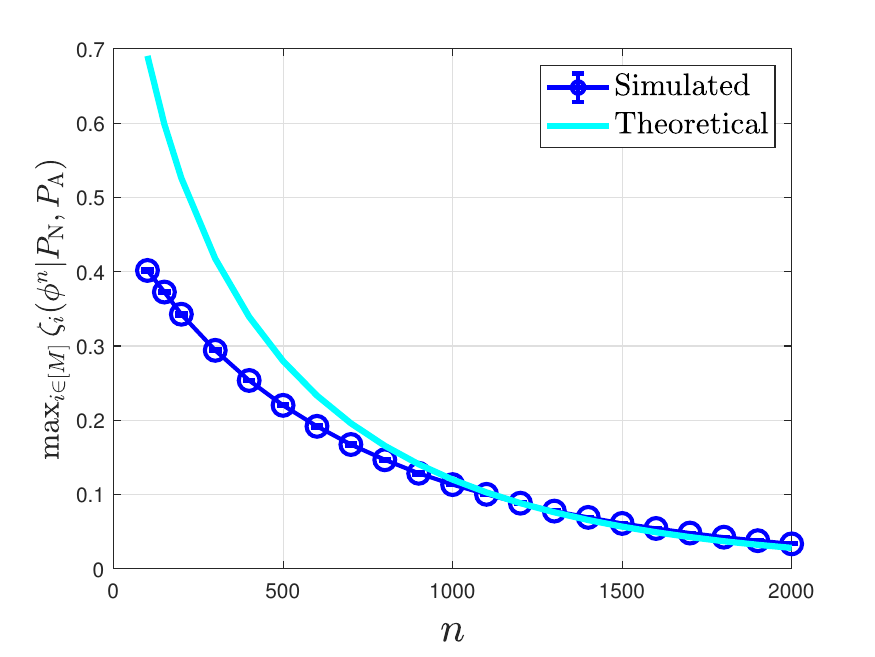}
\caption{Simulated false reject probability when there is one outlying sequence out of $M=4$ sequences. To verify the tightness of our theoretical result, we need to assume a pair of nominal and anomalous distributions to permit the calculation of the false reject probability. For this purpose, the nominal distribution is assumed to be $P_\rmN=\mathrm{Bern}(0.2)$ and the anomalous distribution is assumed to be $P_\rmA=\mathrm{Bern}(0.4)$. The error bar denotes one standard deviation below and above the mean value. As observed, the simulated false reject probability approaches the target value $\varepsilon$ as the lengths of observed sequences become moderate. This implies that our theoretical upper bound on the false reject probability in Theorem \ref{result:1outlier} is tight for this numerical example.}
\label{explain_theorem}
\end{figure}

We next simulate the false reject probability of our test in \eqref{test1outlier} with $\lambda=0.05$
when the nominal distribution is $P_\rmN=\mathrm{Bern}(0.2)$ and the anomalous distribution is $P_\rmA=\mathrm{Bern}(0.4)$. The simulation results are plotted and compared with the theoretical upper bound in Figure \ref{explain_theorem}. Specifically, the theoretical result corresponds to the upper bound in \eqref{upp3} where the $O(\frac{\log n}{n})$ term is chosen as $\frac{\log n}{n}$ and the additive term $O(\frac{1}{\sqrt{n}})$ is ignored. From Figure \ref{explain_theorem}, we find that our theoretical upper bound on the false reject probability in Theorem \ref{result:1outlier} is rather tight for $n\geq 1000$ in this numerical example.

\begin{figure}[tb]
\centering
\includegraphics[width=.5\columnwidth]{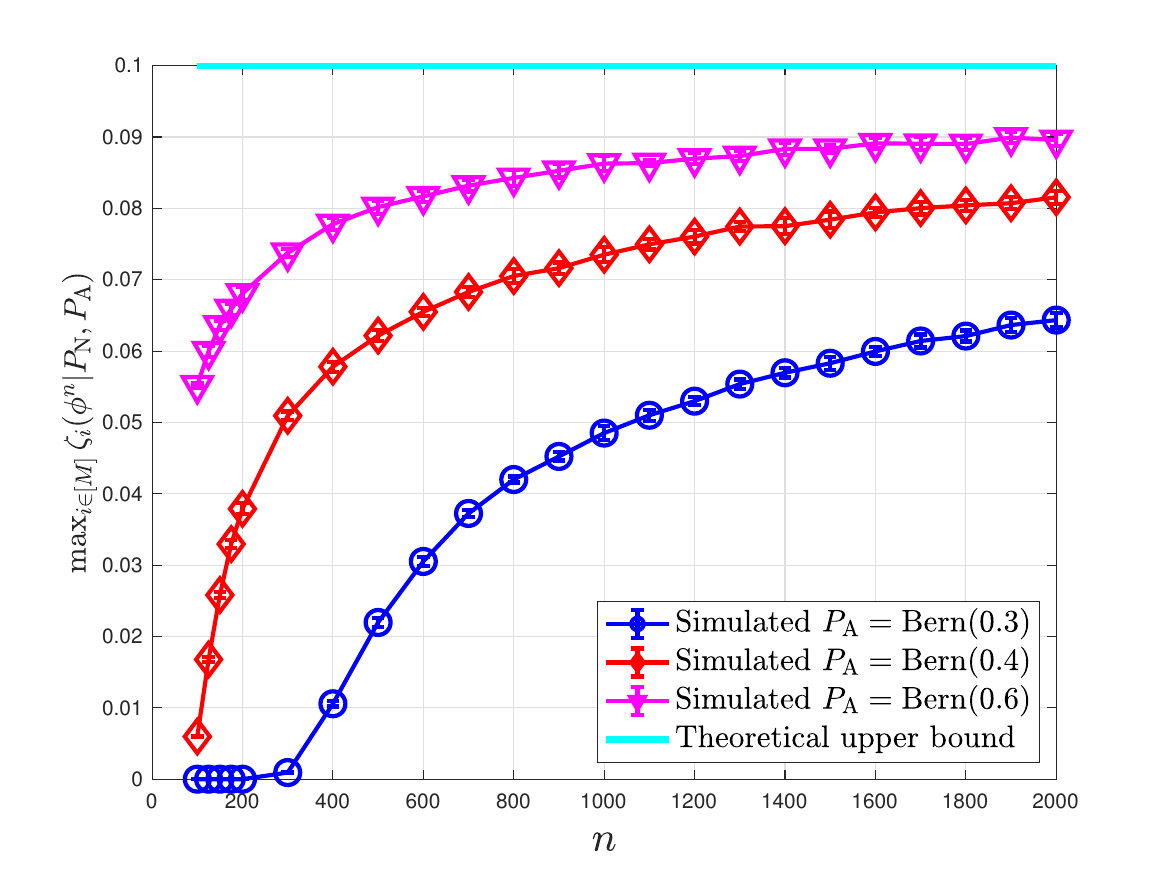}
\caption{Simulated false reject probability when there is one outlying sequence out of $M=4$ sequences to illustrate Corollary \ref{coro:1outlier}. For this purpose, the nominal distribution is assumed to be $P_\rmN=\mathrm{Bern}(0.2)$ and various anomalous distributions are considered. The error bar denotes one standard deviation below and above the mean value. As observed, the simulated false reject probability approaches the target value $\varepsilon$ as the lengths of observed sequences become moderate, as predicted in Corollary \ref{coro:1outlier}.}
\label{numericalSim}
\end{figure}

Finally, to illustrate Corollary \ref{coro:1outlier}, we further simulate the false reject probability of the test in \eqref{test1outlier} with the following threshold
\begin{align}
\lambda_n=\mathrm{GD}_M(P_\rmN,P_\rmA)+\frac{L_M^*(\varepsilon|P_\rmN,P_\rmA)}{\sqrt{n}},
\end{align}
for $\varepsilon=0.1$ and $n\in\{100,125,\ldots,200,300,\ldots,1500\}$. Corollary \ref{coro:1outlier} claims that the false reject probability of our test in \eqref{test1outlier} is upper bound by $\varepsilon=0.1$ asymptotically. We plot the simulated results versus the theoretical upper bound in Figure \ref{numericalSim} for the nominal distribution $P_\rmN=\mathrm{Bern}(0.2)$ and different anomalous distributions $P_\rmA$. From Figure \ref{numericalSim}, we find that for all cases, the simulated false reject probability approaches $\varepsilon=0.1$ as $n$ increases. The gap between the simulated result and the theoretical upper bound results from the uncharacterized third-order term $O(\frac{\log n}{n})$ in Theorem \ref{result:1outlier}. We remark that the simulated false reject probability for $P_\rmA=\mathrm{Bern}(0.6)$ is closer to the target value $\varepsilon=0.1$ than the other cases because the uncharacterized third-order term $O(\frac{\log n}{n})$ has relatively smaller influence for a larger $\rmG_M(P_\rmN,P_\rmA)$.

\subsection{Asymptotic Decay Rates}
For accurate anomaly detection, all error probabilities should be small to ensure that no outlying sequence is missed or identified incorrectly. Thus, a constant or even vanishing false reject probability might not suffice when the length of the observed sequence $n$ is unbounded. In the following theorem, we obtain an asymptotic tradeoff between the exponents of false reject probabilities and the homogeneous error exponent for misclassification error and false alarm probabilities.
Recall the definition of $\mathrm{LD}_i(\lambda,P_\rmN,P_\rmA)$ in \eqref{def:LDi}.

\begin{theorem}
\label{result:exp1out}
For every pair of nominal and anomalous distributions $(P_\rmN,P_\rmA)$, given any positive real number $\lambda\in\bbR_+$, the test in \eqref{test1outlier} satisfies:
\begin{align}
\liminf_{n\to\infty}\frac{1}{n}\min\left\{\min_{i\in[M]}\{-\log\beta_i(\psi_n|P_\rmN,P_\rmA)\},-\log\rmP_{\mathrm{fa}}(\psi_n|P_\rmN,P_\rmA)\right\}&\geq \lambda,\\
\liminf_{n\to\infty}-\frac{1}{n}\log\zeta_i(\psi_n|P_\rmN,P_\rmA)&\geq \mathrm{LD}_i(\lambda,P_\rmN,P_\rmA).
\end{align}
Conversely, given any positive real number $\lambda\in\bbR_+$, for any test $\phi_n$ such that for all pairs of nominal distributions $(\tilP_\rmN,\tilP_\rmA)$,
\begin{align}           
\liminf_{n\to\infty}\frac{1}{n}\min\left\{\min_{i\in[M]}\{-\log\beta_i(\phi_n|\tilP_\rmN,\tilP_\rmA)\},-\log\rmP_{\mathrm{fa}}(\psi_n|\tilP_\rmN,\tilP_\rmA)\right\}&\geq \lambda\label{concondition},
\end{align}
under any pair of nominal and anomalous distributions $(P_\rmN,P_\rmA)$, the false reject exponent satisfies
\begin{align}
\limsup_{n\to\infty}-\frac{1}{n}\log\zeta_i(\phi_n|P_\rmN,P_\rmA)&\leq \mathrm{LD}_i(\lambda,P_\rmN,P_\rmA).
\end{align}

\end{theorem}
The differences between the proof of Theorem \ref{result:exp1out} and the proofs of Theorems \ref{result:1outlier}, \ref{converse:1outlier} lie in the analysis of the false reject probability. See Appendix \ref{proof:result:exp1out}.

To ensure that all three kinds of error probabilities decay exponentially, we need $\min\{\lambda,\min_{i\in[M]}\mathrm{LD}_i(\lambda,P_\rmN,P_\rmA)\}>0$. Given any $(P_\rmN,P_\rmA)$, for each $i\in[M]$, $\mathrm{LD}_i(\lambda,P_\rmN,P_\rmA)$ (cf. \eqref{def:LDi}) is non-increasing in $\lambda$ and $\mathrm{LD}_i(\lambda,P_\rmN,P_\rmA)=0$ if and only if $\lambda\geq \mathrm{GD}_M(P_N,P_A)$ (cf. Appendix \ref{sec:just} for justification). Therefore, for any pair of nominal and anomalous distributions $(P_\rmN,P_\rmA)$ such that $\mathrm{GD}_M(P_\rmN,P_\rmA)>0$ and $\lambda<\mathrm{GD}_M(P_\rmN,P_\rmA)$, all three kinds of error probabilities decay to zero exponentially fast. However, in practice, one cannot know either $P_\rmN$ or $P_\rmA$. Thus, the above result implies that one can choose a smaller threshold $\lambda$ to ensure exponentially consistent performance under a larger set of distributions. When one has some information about the underling true pair of distributions $(P_\rmN,P_\rmA)$, one can choose a large enough $\lambda$ to ensure good homogeneous error exponent and a positive false reject exponent.

We further discuss the tradeoff between the false reject exponent $\mathrm{LD}_i(\lambda,P_\rmN,P_\rmA)$ and the homogeneous error exponent $\lambda$ under each hypothesis. Specifically, one might wonder what value is taken on by the largest false reject exponent given any positive $\lambda$. In Appendix \ref{sec:just}, we show that
\begin{align}
\sup_{\lambda\in\bbR_+}\mathrm{LD}_i(\lambda,P_\rmN,P_\rmA)<\min_{Q\in\calP(\calX)}(D(Q\|P_\rmA)+(M-1)D(Q\|P_\rmN))\label{toprove2},
\end{align}
and thus provide an answer to the above question. Note that the right hand side in \eqref{toprove2} is positive if $P_\rmN\neq P_\rmA$.

The converse part states that the test in \eqref{test1outlier} is also optimal under the generalized Neyman-Pearson criterion when the false reject probabilities decay exponentially fast. Specifically, among all tests that ensure exponential decay of misclassification error and false alarm probabilities for all possible pairs of nominal and anomalous distributions, the test in \eqref{test1outlier} has the largest false reject exponent under any pair of nominal and anomalous distributions.

Finally, note that asymptotically the exponents of probabilities of misclassification error and false alarm are equal. This is an artifact of our test in \eqref{test1outlier} where only one threshold $\lambda$ is used. It would be worthwhile to investigate tests that can fully characterize the exponent tradeoff of all three kinds of error probabilities, beyond the degenerate ``corner-point'' case in this paper. Similar comments apply also to our results for the case of multiple outlying sequences. Such investigations will be pursued in future work.

\section{Case of Multiple Outlying Sequences}
\label{sec:Tout}
In this section, we generalize the results in Section \ref{sec:pf} to the case of multiple outlying sequences where each outlying sequence can be generated from a potentially different anomalous distribution. We assume that the number of outlying sequences is unknown but less than half of the total number of the observed sequences. We study the performance of a threshold-based test that generalizes \eqref{test1outlier} and demonstrate the optimality of the test under the generalized Neyman-Pearson criterion.

\subsection{Problem Formulation}

Assume that there are at most $T=:\lceil\frac{M}{2}-1\rceil$ outlying sequences out of $M$ observed sequences $\bX^n=(X_1^n,\ldots,X_M^n)$. In the outlier hypothesis testing problem with at most $T$ outliers, the task is to decide whether there are outlying sequences and identify the set of outlying sequences if any exist. We assume that each outlying sequence is generated i.i.d. from a possibly different anomalous distribution. Specifically, let $\bP_T:=(P_{\rmA,1},\ldots,P_{\rmA,T})$ be a collection of $T$ anomalous distributions that are different from the nominal distribution $P_\rmN$, all defined on the finite alphabet $\calX$ with the same support. Furthermore, for any $t\in[T]$, let $\calS_t$ denote the set of all subsets of $[M]$ whose cardinality (size) is $t$, i.e.,
\begin{align}
\calS_t:=\{\calB\subseteq[M]:~|\calB|=t\}\label{def:calSt}.
\end{align}
Then, define the union of subsets $\calS_t$ over $t\in[T]$ as $\calS:=\bigcup_{t\in[T]}\calS_t$. For any $\calB\in\calS$, let $\bP_\calB$ denote the collection of distributions $(P_{\rmA,1},\ldots,P_{\rmA,|\calB|})$. When $\calB\in\calS$ denotes the index of the outlying sequences, for any $l\in\calB$, $X_l^n$ is generated i.i.d. from $P_{\rmA,\jmath_\calB(l)}$, where $\jmath_\calB$ denotes an ordered mapping from $\calB$ to $[|\calB|]$ such that for each $i\in\calB$, $\jmath_\calB(i):=j$ if $i$ is the $j$-th smallest element in $\calB$. For example, when $M=10$, $\calB=\{2,3,6\}$ and $\bP_\calB=(P_{\rmA,1},P_{\rmA,2},P_{\rmA,3})$, then the second sequence $X_2^n$ is generated i.i.d. from $P_{\rmA,1}$, the third sequence $X_3^n$ is generated i.i.d. from $P_{\rmA,2}$ and the $6$-th sequence $X_6^n$ is generated i.i.d. from $P_{\rmA,3}$ while all other sequences are generated i.i.d. from the unknown nominal distribution $P_\rmN$.

Since the exact number of outlying sequences is \emph{unknown}, there are in total $|\calS|+1=\sum_{t\in[T]}{M \choose t}+1$ possible configurations of outlying sequences. Formally, the task is to design a test $\phi_n:\calX^{Mn}\to \{\{\rmH_\calB\}_{\calB\in\calS},\rmH_\rmr\}$ to classify between the following $|\calS|+1$ hypotheses:
\begin{itemize}
\item $\rmH_\calB$ where $\calB\in\calS$: the set of outlying sequences are sequences $X_j^n$ with $j\in\calB$;
\item $\rmH_\rmr$: there is no outlying sequence.
\end{itemize}
Similarly to Section \ref{sec:pf}, the null hypothesis is introduced to model the case when there is no outlying sequence among all $M$ observed sequences.

Given any test $\phi_n$, under any tuple of nominal and anomalous distributions $(P_\rmN,\bP_T)=(P_\rmN,P_{\rmA,1},\ldots,P_{\rmA,T})$, the performance of  $\phi_n$ is evaluated by the following misclassification error, false reject and false alarm probabilities:
\begin{align}
\beta_{\calB}(\phi_n|P_\rmN,\bP_T)
&:=\bbP_{\calB}\{\phi_n(\bX^n)\notin\{\rmH_\calB,\rmH_\rmr\}\}\label{def:errorsetS},\\
\zeta_{\calB}(\phi_n|P_\rmN,\bP_T)
&:=\bbP_{\calB}\{\phi_n(\bX^n)=\rmH_\rmr\}\label{def:rejectsetS},\\
\rmP_{\mathrm{fa}}(\phi_n|P_\rmN,\bP_T)
&:=\bbP_\rmr\{\phi_n(\bX^n)\neq\rmH_\rmr\}\label{def:falarmM},
\end{align}
where $\calB\in\calS$ denotes the set of indices of outlying sequences, and we define $\bbP_{\calB}(\cdot):=\Pr\{\cdot|\rmH_\calB\}$ where for each $i\in[M]$ such that $i\notin\calB$, $X_i^n$ is generated i.i.d. from the nominal distribution $P_\rmN$ and for $i\in\calB$, $X_i^n$ is generated i.i.d. from an nominal distribution $P_{\rmA,\jmath_\calB(i)}$, finally we define $\bbP_\rmr(\cdot):=\Pr\{\cdot|\rmH_\rmr\}$, where all sequences are generated i.i.d. from the nominal distribution $P_\rmN$. 

\subsection{A Threshold-Based Test}
To present our test, we need the following definition that generalizes $\rmG_i(\bQ)$ in \eqref{def:gi}. Given a sequence of distributions $\bQ=(Q_1,\ldots,Q_M)\in\calP(\calX)^M$ and each $\calB\in\calS$, define the following linear combination of KL divergence terms
\begin{align}
\rmG_\calB(\bQ)
&:=\sum_{t\in\calM_{\calB}}D\left(Q_t\bigg\|\frac{\sum_{l\in\calM_\calB}Q_l}{M-|\calB|}\right)\label{def:gB},
\end{align}
where $\calM_\calB$ is the set of elements that are in $[M]$ but not in $\calB$, i.e., $\calM_\calB:=[M]\setminus\calB=\{i\in[M]:~i\notin\calB\}$. Similar to $\rmG_i(\bQ)$ in \eqref{def:gi}, $\rmG_\calB(\bQ)$ is a homogeneity measure and equals zero if and only if $Q_j=Q$ for all $j\in\calM_\calB$ where $Q\in\calP(\calX)$ is arbitrary.

Throughout the section, we use a threshold-based test that takes the empirical distribution of each observed sequence as the input and outputs a decision among all hypotheses. Given $M$ observed sequences $\bx^n=(x_1^n,\ldots,x_M^n)$ and any positive real number $\lambda$, the test operates as follows:
\begin{align}
\label{test:Toutlier}
\Psi_n(\bx^n)
&:=\left\{
\begin{array}{cc}
\rmH_\calB&\mathrm{if~}\rmS_\calB(\bx^n)<\min_{\calC\in\calS_\calB}\rmS_\calC(\bx^n)\mathrm{~and~}\min_{\calC\in\calS_\calB}\rmS_\calC(\bx^n)>\lambda,\\
\rmH_\rmr&\mathrm{otherwise},
\end{array}
\right.
\end{align}
where $\calS_\calB=\{\calC\in\calS_\calB\}$ and $\rmS_\calC(\cdot)$ is the scoring function defined as
\begin{align}
\rmS_\calC(\bx^n)
&:=\rmG_\calC(\hatT_{x_1^n},\ldots,\hatT_{x_M^n})\label{def:scalC},
\end{align}
which measures the sum of KL divergence between the empirical distribution of each sequence $x_j^n$ with $j\notin\calB$ relative to the average of the empirical distributions of all sequences $x_j^n$ where $j\notin\calB$. For the special case of $T=1$, the test in \eqref{test:Toutlier} reduces to the test in \eqref{test1outlier}.

We then discuss how the test in \eqref{test:Toutlier} deals with different and unknown number of outlying sequences when $T\geq 2$. Given $M$ observed sequences $\bx^n$, we calculate the scoring functions $\rmG_\calB(\hatT_{x_1^n},\ldots,\hatT_{x_M^n})$ for all possible sets $\calB\subseteq\calS$. Note that each $\calB\subseteq(\calS\setminus\emptyset)$ denotes a possible set of indices of outlying sequences and $\calB=\emptyset$ corresponds to the null hypothesis that no outlying sequence appears. To determine the set of outlying sequences, using the scoring function for all possible ${M\choose T}+1$ cases, we run the test in \eqref{test:Toutlier} that compares each scoring function with the threshold $\lambda$. In other words, the test \eqref{test:Toutlier} checks all possibilities of outlying sequences to make a decision and its complexity increases exponentially with $T$. Note that the test in \eqref{test:Toutlier} is a generalization of our test in \eqref{test1outlier} for the case of at most one outlying sequence and specializes to \eqref{test1outlier} when $T=1$. The property of test in \eqref{test:Toutlier} is similar to the discussion of the test in \eqref{test1outlier}.

Finally, we remark that the statistic in \eqref{def:gB} was also used in \cite[Eq. (37)]{li2014} to construct a test when the number $t$ of outlying sequences is known and when there is no null hypothesis. In contrast, the test in \eqref{test:Toutlier} does not assume any knowledge of the number of outlying sequences, and in addition, incorporates a null hypothesis to include the possibility of no outliers. 

\subsection{Preliminaries}
To present our main results, we need the following definitions that generalize those in Section \ref{sec:prelim} for the case of at most one outlying sequence. Given any $\calB\in\calS$ and any tuple of distributions $\bP_\calB=(P_\rmN,P_{\rmA,1},\ldots,P_{\rmA,|\calB|})\in(\calP(\calX))^{|\calB|+1}$, for any two sets $(\calB,\calC)\in\calS^2$, define the following mixture distribution
\begin{align}
P_{\rm{Mix}}^{(\calB,\calC,P_\rmN,\bP_\calB)}(x)
&:=\frac{1}{M-|\calC|}\Big(\sum_{i\in(\calB\cap\calM_{\calC})}P_{\rmA,\jmath_\calB(i)}(x)+\sum_{i\in(\calM_{\calB}\cap\calM_{\calC})}P_\rmN(x)\Big)\label{def:Pmix},
\end{align}
and, parallel to \eqref{def:id1} and \eqref{def:id2}, define the following information densities (log likelihoods)
\begin{align}
\imath_{1,l}(x|\calB,\calC,P_\rmN,\bP_\calB)&:=\log\frac{P_{\rmA,l}(x)}{P_{\rm{Mix}}^{(\calB,\calC,P_\rmN,\bP_\calB)}(x)},~l\in[|\calB|]\label{def:i1T},\\
\imath_2(x|\calB,\calC,P_\rmN,\bP_\calB)&:=\log\frac{P_\rmN(x)}{P_{\rm{Mix}}^{(\calB,\calC,P_\rmN,\bP_\calB)}(x)}\label{def:i2T}.
\end{align}

Analogously to \eqref{def:GD} to \eqref{def:cov}, define the following linear combinations of expectations and variances of information densities:
\begin{align}
\nn&\mathrm{GD}(\calB,\calC,P_\rmN,\bP_\calB)\\*
&:=\sum_{i\in(\calB\cap\calM_{\calC})}\mathbb{E}_{P_{\rmA,\jmath_\calB(i)}}[\imath_{1,\jmath_\calB(i)}(X|\calB,\calC,P_\rmN,\bP_\calB)]+\sum_{i\in(\calM_{\calB}\cap\calM_{\calC})}\mathbb{E}_{P_\rmN}[\imath_2(X|\calB,\calC,P_\rmN,\bP_\calB)]\\
&=\sum_{i\in(\calB\cap\calM_{\calC})}D(P_{\rmA,\jmath_\calB(i)}\|P_{\rm{Mix}}^{(\calB,\calC,P_\rmN,\bP_\calB)})+\sum_{i\in(\calM_{\calB}\cap\calM_{\calC})}D(P_{\rmN}\|P_{\rm{Mix}}^{(\calB,\calC,P_\rmN,\bP_\calB)})\label{def:GDBC},\\
\nn&\rmV(\calB,\calC,P_\rmN,\bP_\calB)\\*
&:=\sum_{i\in(\calB\cap\calM_{\calC})}\mathrm{Var}_{P_{\rmA,\jmath_\calB(i)}}[\imath_{1,\jmath_\calB(i)}(X|\calB,\calC,P_\rmN,\bP_\calB)]+\sum_{i\in(\calM_{\calB}\cap\calM_{\calC})}\mathrm{Var}_{P_\rmN}[\imath_2(X|\calB,\calC,P_\rmN,\bP_\calB)]\label{def:VBC}.
\end{align}

For simplicity, given any $(\calB,\calC)\in\calS^2$ and any variables $(x_1,\ldots,x_M)$, let
\begin{align}
\imath_{\calB,\calC}(x_1,\ldots,x_M|P_\rmN,\bP_\calB)
&:=\sum_{j\in(\calB\cap\calM_\calC)}\imath_{1,\jmath_\calB(j)}(x_j|\calB,\calC,P_\rmN,\bP_\calB)+\sum_{\barj\in(\calM_\calB\cap\calM_\calC)}\imath_2(x_{\barj}|\calB,\calC|P_\rmN,\bP_\calB).
\end{align}
For ease of latter presentation, let $\calS_\calB$ denote the set $\calS\setminus\{\calB\}$, i.e., $\{\calC\in\calS:\calC\neq \calB\}$. Furthermore, let the elements in $\calS_\calB$ be ordered as $\{\calC_1,\ldots,\calC_{|\calS|-1}\}$. Then for each $(i,k)\in[|\calS|-1]^2$ such that $i\neq k$, define the covariance
\begin{align}
\mathrm{Cov}(\calC_i,\calC_k,P_\rmN,\bP_\calB)
\nn&:=\mathrm{E}[\imath_{\calB,\calC_i}(X_1,\ldots,X_M|P_\rmN,\bP_\calB)\imath_{\calB,\calC_k}(X_1,\ldots,X_M|P_\rmN,\bP_\calB)].
\end{align}
Analogously to \eqref{defdispersionmatrix}, define a covariance matrix $\bV(\calB,P_\rmN,\bP_\calB)=\{V_{i,j}(\calB,P_\rmN,\bP_\calB)\}_{(i,j)\in[|\calS|-1]^2}$ where
\begin{align}
V_{i,j}(\calB,P_\rmN,\bP_\calB)
&=\left\{
\begin{array}{ll}
\rmV(\calB,\calC_i,P_\rmN,\bP_\calB)&\mathrm{if~}i=j,\\
\mathrm{Cov}(\calC_i,\calC_k,P_\rmN,\bP_\calB)&\mathrm{otherwise.}
\end{array}
\right.
\end{align}
The complementary cdf $\rmQ_k(\cdot)$ in \eqref{def:kQ}, together with $\mathrm{GD}(\calB,\calC,P_\rmN,\bP_\calB)$ and $\bV(\calB,P_\rmN,\bP_\calB)$, will be critical to upper bound the false reject probabilities.

Finally, given any $\lambda\in\bbR_+$ and any tuple of distributions $\bP_\calB=(P_\rmN,P_{\rmA,1},\ldots,P_{\rmA,T})\in\calP_T(\calX)$, for each $\calB\in\calS$, define the following quantity:
\begin{align}
\mathrm{LD}_{\calB}(\lambda,P_\rmN,\bP_\calB)
&:=\min_{(\calC,\calD)\in\calS^2:\calC\neq\calD}\min_{\substack{\bQ\in(\calP(\calX))^M:\\\rmG_\calC(\bQ)\leq \lambda,~\rmG_\calD(\bQ)\leq\lambda}}\Big(\sum_{i\in\calB}D(Q_i\|P_{\rmA,\jmath_\calB(i)})+\sum_{i\in\calM_\calB}D(Q_i\|P_\rmN)\Big)\label{def:LDT}.
\end{align}
The quantity $\mathrm{LD}_{\calB}(\lambda,P_\rmN,\bP_\calB)$ will characterize the false reject exponent under each hypothesis.

\subsection{Second-Order Asymptotic Approximation to the Non-Asymptotic Performance}
Our first set of results characterize the performance tradeoff among probabilities of misclassification error, false alarm and false reject. Specifically, we first provide an achievability result, where the performance of the test in \eqref{test:Toutlier} is characterized in terms of misclassification error and false alarm probabilities that decay exponentially fast when the false reject probability is upper bounded by a function of the threshold $\lambda$. Furthermore, we demonstrate the optimality of the test in \eqref{test:Toutlier} under the generalized Neyman-Pearson criterion.

\begin{theorem}
\label{result:Toutlier}
For any nominal distribution $P_\rmN$ and anomalous distributions $\bP_T=(P_{\rmA,1},\ldots,P_{\rmA,T})$, given any positive real number $\lambda\in\bbR_+$, the test in \eqref{test:Toutlier} satisfies that for each $\calB\in\calS$,
\begin{align}
\beta_{\calB}(\Psi_n|P_\rmN,\bP_T)&\leq \exp\Big(-n\lambda+|\calX|\log((M-1)n+1)\Big),\\
\rmP_{\mathrm{fa}}(\Psi_n|P_\rmN,\bP_T)&\leq |\calS|^2\exp\Big(-n\lambda+|\calX|\log((M-1)n+1)\Big),\\
\zeta_\calB(\Psi_n|P_\rmN,\bP_T)&\leq 1-\rmQ_{|\calS|-1}\big(\sqrt{n}\bar{\mu}(\lambda,P_\rmN,\bP_\calB);\mathbf{0}_{|\calS|-1};\bV(\calB,P_\rmN,\bP_\calB)\big)+O\left(\frac{1}{\sqrt{n}}\right)\label{sr:tout},
\end{align} 
where $\bar{\mu}(\lambda,P_\rmN,\bP_\calB)$ denotes the vector $(\lambda-\mathrm{GD}(\calB,\calC_1,P_\rmN,\bP_\calB)+O(\log n/n),\ldots,\lambda-\mathrm{GD}(\calB,\calC_{|\calS|-1},P_\rmN,\bP_\calB)+O(\log n/n))$.
\end{theorem}
The proof of Theorem \ref{result:Toutlier} is a generalization of the proof of Theorem \ref{result:1outlier} and is given in Appendix \ref{proof:result:Toutlier}.

Similarly to the result in Theorem \ref{result:1outlier}, when the number of outlying sequences $M$ is finite, both misclassification error  false alarm probabilities decay exponentially fast, with a speed lower bounded by $\lambda$ asymptotically when $n$ tends to infinity. On the other hand, the false reject under each hypothesis $\rmH_\calB$ is upper bounded by a function of $\lambda$ and critical quantities $\mathrm{GD}(\calB,\calC,P_\rmN,\bP_\calB)$ and $\bV(\calB,P_\rmN,\bP_\calB)$. Note that the threshold $\lambda$ trades off the lower bound on the decay rate of the homogeneous error exponent of the misclassification error and false alarm probabilities and the upper bound on the false reject probability. If $\lambda$ increases, the homogeneous error exponent increases while the false reject probability increases as well. This implies that better performance in misclassification error and false alarm probabilities leads to worse false reject probabilities.

Asymptotically as $n\to\infty$, if the threshold $\lambda<\min_{i\in[|\calS|-1]}\mathrm{GD}(\calB,\calC_i,P_\rmN,\bP_\calB)$, then the false reject probability under hypothesis $\rmH_\calB$ vanishes. One might also be interested in the more practical non-asymptotic case where $n$ is finite. Obtaining the exact solution to such case is almost impossible. However, a second-order asymptotic approximation to the non-asymptotic performance is possible using the result in \eqref{sr:tout}. For this purpose, we define
\begin{align}
\mathrm{GD}(\calB,P_\rmN,\bP_{\calB})
:=\min_{i\in[|\calS|-1]}\mathrm{GD}(\calB,\calC_i,P_\rmN,\bP_\calB)
\end{align}
as the minimum value of the vector $(\mathrm{GD}(\calB,\calC_1,P_\rmN,\bP_\calB),\ldots,\mathrm{GD}(\calB,\calC_{|\calS|-1},P_\rmN,\bP_\calB))$ and let $d(\calB)$ be the number of elements in the vector that equals the minimal value, i.e., $d(\calB):=\big|\{i\in[|\calS|-1]:\mathrm{GD}(\calB,\calC_i,P_\rmN,\bP_\calB)=\mathrm{GD}(\calB,P_\rmN,\bP_{\calB})\}\big|$. Analogously to \eqref{def:L*PNPA} and \eqref{def:lambda^*}, given any $\varepsilon\in(0,1)$, let
\begin{align}
L^*(\varepsilon|\calB,P_\rmN,\bP_\calB)
&:=\max\Big\{L\in\bbR:\rmQ_{d(\calB)}(L\times\mathbf{1}_{d(\calB)};\mathbf{0}_{d(\calB)};\bV(\calB,P_\rmN,\bP_\calB))\geq 1-\varepsilon\Big\}\label{def:L*PT},\\
\lambda^*(n,\varepsilon|\calB,P_\rmN,\bP_\calB)&:=\mathrm{GD}(\calB,P_\rmN,\bP_{\calB})+\frac{L^*(\varepsilon|\calB,P_\rmN,\bP_\calB)}{\sqrt{n}}\label{def:lambda^*moutlier}.
\end{align}
We then have the following corollary of Theorem \ref{result:Toutlier}.
\begin{corollary}
\label{second:tout}
For any $(\calB,P_\rmN,\bP_\calB)$, if $\lambda$ satisfies $\lambda\leq \lambda^*(n,\varepsilon|\calB,P_\rmN,\bP_\calB)$ for all $n\in\bbN$, then as $n$ increases, the upper bound on the false reject probability tends to $\varepsilon\in(0,1)$, i.e., $\limsup_{n\to\infty}\zeta_\calB(\Psi_n|P_\rmN,\bP_T)\leq \varepsilon$.
\end{corollary}

The second-order asymptotic upper bound in Corollary \ref{second:tout} provides further characterization beyond the first-order asymptotic constant term $\mathrm{GD}(\calB,P_\rmN,\bP_{\calB})$ and it trades off the homogeneous error exponent with any any non-vanishing false reject probability $\varepsilon\in(0,1)$ beyond the vanishing case with $\varepsilon\to 0$ implied by a first-order asymptotic analysis. 

Finally, we discuss the influence of the number of observed sequences $M$ on the performance of the test in \eqref{test:Toutlier}. As demonstrated in the above remark, $\mathrm{GD}(\calB,P_\rmN,\bP_{\calB})$ is the critical quantity that is related with the performance of the test. Thus, it suffices to study the properties of $\mathrm{GD}(\calB,P_\rmN,\bP_{\calB})$ as a function of $M$ under each hypothesis $\rmH_\calB$. However, it is challenging to obtain closed form equations for the dependence of $\mathrm{GD}(\calB,P_\rmN,\bP_{\calB})$ on $M$ when each outlying sequence is generated from a unique anomalous distributions. Thus, we specialize our results to the case where all anomalous distributions are the same and denoted by $P_\rmA$. Under this assumption, one can verify that
\begin{align}
\mathrm{GD}(\calB,P_\rmN,\bP_{\calB})
&=\min_{t\in[T]}\min_{l\in[|\calB|]}\big(lD(P_\rmA\|P_{\mathrm{Mix}}^{t,l})+(M-t-l)D(P_\rmN\|P_{\mathrm{Mix}}^{l,t})\big)\label{allsame},
\end{align}
where $P_{\mathrm{Mix}}^{t,l}=\frac{lP_\rmA+(M-t-l)P_\rmN}{M-t}$. For any $(t,l)\in[T]\times[|\calB|]$, one can verify that
\begin{align}
\frac{\partial \mathrm{GD}(\calB,P_\rmN,\bP_{\calB})}{\partial M}
&=D(P_\rmN\|P_{\mathrm{Mix}}^{l,t}).
\end{align}
Thus, $\mathrm{GD}(\calB,P_\rmN,\bP_{\calB})$ increases in $M$ if $D(P_\rmN\|P_{\mathrm{Mix}}^{l,t})>0$, which holds for all distinct pair of nominal and anomalous distributions. This implies that the performance of the test in \eqref{test:Toutlier} increases as the number of observed sequences $M$ increases when the number of outlying sequences $|\calB|$ remains unchanged. On the other hand, the result in \eqref{allsame} implies that for a fixed number of observed sequences $M$, the performance of the test in \eqref{test:Toutlier} degrades as the number of outlying sequences $|\calB|$ increases.

In the following theorem, it is shown that the test in \eqref{test:Toutlier} is optimal under the generalized Neyman-Pearson criterion for second-order asymptotic analysis.
\begin{theorem}
\label{converse:m}
Given any $\lambda\in\bbR_+$, for any test $\phi_n$ such that
\begin{align}
\beta_\calB(\phi_n|\tilP_\rmN,\tilde{\bP}_T)\leq \exp(-n\lambda),~\forall~(\tilP_\rmN,\tilde{\bP}_T),
\end{align}
then for any tuple of nominal and anomalous distributions $(P_\rmN,\bP_T)$, for each $\calB\in\calS$, 
\begin{align}
\zeta_\calB(\Psi_n|P_\rmN,\bP_T)\geq 1-\rmQ_{|\calS|-1}\big(\sqrt{n}\bar{\mu}(\lambda,P_\rmN,\bP_\calB);\mathbf{0}_{|\calS|-1};\bV(\calB,P_\rmN,\bP_\calB)\big)+O\left(\frac{1}{\sqrt{n}}\right).
\end{align}
\end{theorem}
The proof of Theorem \ref{converse:m} is similar to that of Theorem \ref{converse:1outlier} and only salient differences are emphasized in Appendix \ref{converse:toutlier}.

\subsection{Asymptotic Decay Rates}
We next study the case where the false reject probability decays exponentially fast as well and thus characterize the tradeoff between the false reject exponent and the homogeneous error exponent of the misclassification error and false alarm probabilities. Recall the definition of $\mathrm{LD}_{\calB}(\lambda,P_\rmN,\bP_\calB)$ in \eqref{def:LDT}.
\begin{theorem}
\label{result:Tout:exp}
For any nominal distribution $P_\rmN$ and anomalous distributions $\bP_T=(P_{\rmA,1},\ldots,P_{\rmA,T})$, given any positive real number $\lambda\in\bbR_+$, the test in \eqref{test:Toutlier} satisfies that for each $\calB\in\calS$,
\begin{align}
\liminf_{n\to\infty}-\frac{1}{n}\log\beta_{\calB}(\Psi_n|P_\rmN,\bP_T)&\geq\lambda,\\
\liminf_{n\to\infty}-\frac{1}{n}\log\rmP_{\mathrm{fa}}(\Psi_n|P_\rmN,\bP_T)&\geq \lambda,\\
\liminf_{n\to\infty}-\frac{1}{n}\log\zeta_\calB(\Psi_n|P_\rmN,\bP_T)&\geq\mathrm{LD}_{\calB}(\lambda,P_\rmN,\bP_\calB).
\end{align} 
Conversely, for any test that ensures the homogeneous exponential decay rate of the misclassification error and false alarm is no less than $\lambda$ for all tuples of nominal and anomalous distributions, under any nominal distribution $P_\rmN$ and anomalous distributions $\bP_T=(P_{\rmA,1},\ldots,P_{\rmA,T})$, the false reject exponent is also upper bounded by $\mathrm{LD}_{\calB}(\lambda,P_\rmN,\bP_\calB)$ under each hypothesis $\rmH_\calB$.
\end{theorem}
The proof of Theorem \ref{result:Tout:exp} is omitted since it requires modifying the proof of Theorem \ref{result:Toutlier} similarly to how one modifies the proof of Theorem \ref{result:1outlier} to prove Theorem \ref{result:exp1out}. The result in Theorem \ref{result:exp1out} follows by specializing Theorem \ref{result:Tout:exp} to the case of $T=1$. Similar remarks as those for Theorem \ref{result:exp1out} apply here. 

For example, the threshold $\lambda$ governs the tradeoff between the false reject exponent and the homogeneous error exponent under each hypothesis. From the definition of $\mathrm{LD}_{\calB}(\lambda,P_\rmN,\bP_\calB)$ in \eqref{def:LDT}, it follows that the false reject exponent $\mathrm{LD}_{\calB}(\lambda,P_\rmN,\bP_\calB)$ in \eqref{def:LDT} decreases in $\lambda$. Similarly to the proof in Appendix \ref{sec:just}, one can show that $\mathrm{LD}_{\calB}(\lambda,P_\rmN,\bP_\calB)>0$ if and only if $\lambda<\mathrm{GD}(\calB,P_\rmN,\bP_\calB)$ and the maximal false reject exponent satisfies
\begin{align}
\max_{\lambda\in(0,\mathrm{GD}(\calB,P_\rmN,\bP_\calB))}\mathrm{LD}_{\calB}(\lambda,P_\rmN,\bP_\calB)
&\leq \min_{Q\in\calP(\calX)} \Big(\sum_{i\in\calB}D(Q\|P_{\rmA,\jmath_{\calB}(i)})+(M-|\calB|)D(Q\|P_\rmN)\Big).
\end{align}
Therefore, if the threshold $\lambda<\min_{\calB\in\calS}\mathrm{GD}(\calB,P_\rmN,\bP_\calB)$, then regardless of the number of outlying sequences, the misclassification error, the false alarm and false reject probabilities decay exponentially fast for any tuple of distributions $(P_\rmN,\bP_T)$ such that $\min_{\calB\in\calS}\mathrm{GD}(\calB,P_\rmN,\bP_\calB)$ is strictly positive.

\section{Conclusion}
\label{sec:conclude}
We revisited the outlier hypothesis testing problem studied by Li \emph{et al.} in \cite{li2014} and derived performance guarantees for tests that are optimal under the generalized Neyman-Pearson criterion~\cite{gutman1989asymptotically}. In particular, we first studied the case with at most one outlying sequence and then generalized our results to the case where there are multiple outlying sequences, the number of outlying sequences is unknown and each outlying sequence can be generated from a unique anomalous distributions. For both cases, we proposed a threshold-based test and analyzed its performance in terms of the tradeoff among the probabilities of misclassification error, false alarm and false reject. Our results have brought new insights beyond \cite{li2014} in several aspects, including the design of a second-order asymptotic optimal test, the dominant factors affecting performance of a test and a second-order asymptotic approximation to the finite sample size performance using finite blocklength information theoretical tools~\cite{polyanskiy2010finite,Tanbook}.

There are several avenues for future research. Firstly, it might be interesting to study tests that can ensure exponential decay of misclassification error probabilities for any pair of nominal and anomalous distributions and simultaneously ensure that the false alarm and false reject probabilities are upper bounded by a constant for all pairs of nominal and anomalous distributions. Secondly, it would be interesting to study the optimality of tests under criteria other than the generalized Neyman-Pearson criterion. For example, whether the tests in this paper are optimal in the finite sample regime for a set of nominal and anomalous distributions, which would be stronger that the asymptotic guarantees provided in this paper. Thirdly, it would be valuable to extend our theory to the scenario where each nominal sample is generated from a different distribution in a neighborhood of a fixed distribution and then derive the performance of the optimal test, similarly to~\cite{hsu2020binary}. Fourthly, one might generalize our results to the case of continuous alphabet where each observed sequence is generated i.i.d. from a probability density function. Finally, it would be worthwhile to consider a sequential setting by incorporating ideas from~\cite{li2017universal} to derive second-order asymptotic limits of an optimal sequential test.

\appendix

\subsection{Proof of Theorem \ref{result:1outlier}}
\label{proof:result:1outlier}

Recall the definitions of information densities in \eqref{def:id1} and \eqref{def:id2}. Given any pair of distributions $(P_\rmN,P_\rmA)$, define the following linear combination of the third absolute moment of information densities
\begin{align}
\rmT(P_\rmN,P_\rmA)
\nn&:=\bbE_{P_\rmA}\Big[\big|\imath_1(X|P_\rmN,P_\rmA)-\bbE_{P_\rmA}[\imath_1(X|P_\rmN,P_\rmA)]\big|^3\Big]\\*
&\qquad+(M-2)\bbE_{P_\rmN}\Big[\big|\imath_2(X|P_\rmN,P_\rmA)-\bbE_{P_\rmN}[\imath_2(X|P_\rmN,P_\rmA)]\big|\Big].
\end{align}
Note that $\rmT(P_\rmN,P_\rmA)$ is finite since we consider distributions $(P_\rmN,P_\rmA)$ with the same support on the finite alphabet $\calX$. Recall the definition of the scoring function $\rmS_i(\bx^n)=\rmG_i(\hatT_{x_1^n},\ldots,\hatT_{x_M^n})$ (cf. \eqref{def:gi}) for each $i\in[M]$. Furthermore, for any given set of sequences $\bx^n=(x_1^n,\ldots,x_M^n)$, define the following two quantities
\begin{align}
i^*(\bx^n)&:=\argmin_{i\in[M]}\rmS_i(\bx^n)\label{def:i*_min},\\
h(\bx^n)&:=\min_{i\in[M]:i\neq i^*(\bx^n)}\rmS_i(\bx^n)\label{def:defh_smin}.
\end{align}
Note that $i^*(\bx^n)$ denotes the index of the minimal scoring function (unique with high probability as we shall show) and $h(\bx^n)$ denotes the value of the second minimal value of the scoring functions. Using these two definitions, our proposed test in \eqref{test1outlier} is equivalently expressed as follows:
\begin{align}
\psi_n(\bx^n)
&=\left\{
\begin{array}{cl}
\rmH_i&\mathrm{if~}i^*(\bx^n)=i,~\mathrm{and~}h(\bx^n)>\lambda,\\
\rmH_\rmr&\mathrm{if~}h(\bx^n)\leq \lambda
\end{array}
\right.
\label{test1:equivalent}
\end{align}

We first analyze the misclassification error probabilities of our test $\psi_n(\cdot)$ under each hypothesis. Recall that we use $\bQ$ to denote a collection of $M$ distributions $(Q_1,\ldots,Q_M)$ defined on the alphabet $\calX$. For any pair of nominal and anomalous distributions $(P_\rmN,P_\rmA)$ and for each $i\in[M]$, we can upper bound the type-$i$ misclassification error probability as follows:
\begin{align}
\nn&\beta_i(\psi_n|P_\rmN,P_\rmA)\\*
&=\bbP_i\{i^*(\bX^n)\neq i,~h(\bX^n)>\lambda\}\\
&\leq\bbP_i\{\rmS_i(\bX^n)>\lambda\}\label{usedefs}\\
&=\sum_{\bx^n\in\calX^{Mn}:\rmS_i(\bx^n)>\lambda}P_\rmA^n(x_i^n)\times \bigg(\prod_{j\in\calM_i}P_\rmN^n(x_j^n)\bigg)\label{usehi}\\
&=\sum_{\substack{\bQ\in(\calP_n(\calX))^M:\\\rmG_i(\bQ)>\lambda}}\quad \sum_{\substack{\bx^n:~\forall~j\in[M]\\ x_j^n\in\calT_{Q_j}^n}}P_\rmA^n(x_i^n)\times \bigg(\prod_{j\in\calM_i}P_\rmN^n(x_j^n)\bigg)\label{usetypes}\\
\nn&=\sum_{\substack{\bQ\in\calP_n(\calX)^M:\\\rmG_i(\bQ)>\lambda}}\quad \sum_{\substack{\bx^n:~\forall~j\in[M]\\ x_j^n\in\calT_{Q_j}^n}}
\exp\bigg(-n\Big(D(Q_i\|P_\rmA)+H(Q_i)\Big)\bigg)\\*
&\qquad\qquad\qquad\qquad\qquad\qquad\times\exp\bigg(-n\Big(\sum_{j\in\calM_i}\big(D(Q_j\|P_\rmN)+H(Q_j)\big)\Big)\bigg)\\
\nn&=\sum_{\substack{\bQ\in\calP_n(\calX)^M:\\\rmG_i(\bQ)>\lambda}}\quad\sum_{\substack{\bx^n:~\forall~j\in[M]\\ x_j^n\in\calT_{Q_j}^n}}
\exp\bigg(-n\Big(\sum_{t\in[M]}H(Q_t)+D(Q_i\|P_\rmA)\Big)\bigg)\\*
&\qquad\qquad\qquad\qquad\times
\exp\bigg(-n\bigg(\rmG_i(\bQ)+(M-1)D\bigg(\frac{\sum_{k\in\calM_i}Q_k}{M-1}\bigg\|P_\rmN\bigg)\bigg)\bigg)\label{useeqn}\\
\nn&\leq\exp(-n\lambda)\sum_{\substack{\bQ\in\calP_n(\calX)^M}}\sum_{x_i^n\in\calT_{Q_i}^n}
\exp\bigg(-n\Big(D(Q_i\|P_\rmA)+H(Q_i)\Big)\bigg)\\*
&\qquad\qquad\qquad\qquad\times
\exp\bigg(-n\bigg((M-1)D\bigg(\frac{\sum_{k\in\calM_i}Q_k}{M-1}\bigg\|P_\rmN\bigg)\bigg)\bigg)\label{usetypesize}\\
&\leq\exp(-n\lambda)\sum_{Q_j\in\calP_n(\calX),~j\in\calM_i}\exp\bigg(-n\bigg((M-1)D\bigg(\frac{\sum_{k\in\calM_i}Q_k}{M-1}\bigg\|P_\rmN\bigg)\bigg)\bigg)\label{sumxi}\\
&\leq\exp(-n\lambda)\sum_{Q\in\calP^{(M-1)n}(\calX)}((M-1)n+1)^{|\calX|}P_\rmN^{(M-1)n}\Big(\calT_Q^{(M-1)n}\Big)\label{concat}\\
&=\exp\Big(-n\lambda+|\calX|\log((M-1)n+1)\Big)\label{torefer},
\end{align}
where \eqref{usedefs} follows from definitions of $i^*(\bx^n)$ in \eqref{def:i*_min} and $h(\bx^n)$ in \eqref{def:defh_smin} which indicate that $\rmS_i(\bx^n)\geq h(\bx^n)>\lambda$ under the condition that $i^*(\bx^n)\neq i$ and $h(\bx^n)>\lambda$; \eqref{usehi} follows since under hypothesis $\rmH_i$, the $i$-th sequence $X_i^n$ is generated i.i.d. according to the anomalous distribution $P_\rmA$ while all other sequences are generated i.i.d. according to the nominal distribution $P_\rmN$; \eqref{usetypes} follows from the definitions of the scoring function $\rmS_i(\cdot)$ in \eqref{def:scoref} and $\rmG_i(\cdot)$ in \eqref{def:gi} and method of types~\cite[Chapter 11]{cover2012elements}; \eqref{useeqn} follows since for any sequence of distributions $\bQ=(Q_1,\ldots,Q_M)$ and any distribution $\tilP_\rmN$, the following equalities hold
\begin{align}
\sum_{j\in\calM_i}D(Q_j\|P_\rmN)
&=\sum_{j\in\calM_i}\mathbb{E}_{Q_j}\left[\log\frac{Q_j(X)}{P_\rmN(X)}\right]\\
&=\sum_{j\in\calM_i}\mathbb{E}_{Q_j}\left[\log\frac{\frac{1}{M-1}\sum_{k\in\calM_i}Q_k(X)}{P_\rmN(X)}+\log\frac{Q_j(X)}{\frac{1}{M-1}\sum_{k\in\calM_i}Q_k(X)}\right]\\
&=\sum_{j\in\calM_i}\mathbb{E}_{Q_j}\left[\log\frac{\frac{1}{M-1}\sum_{k\in\calM_i}Q_k(X)}{P_\rmN(X)}\right]+\rmG_i(\bQ)\\
&=(M-1)D\bigg(\frac{\sum_{k\in\calM_i}Q_k}{M-1}\bigg\|P_\rmN\bigg)+\rmG_i(\bQ);
\end{align}
\eqref{usetypesize} follows since the size of the type class $|\calT_{Q_j}^n|\leq \exp(nH(Q_j))$; \eqref{sumxi} follows since
\begin{align}
\sum_{Q_i\in\calP_n(\calX)}\sum_{x_i^n\in\calT_{Q_i}^n}\exp\bigg(-n\Big(D(Q_i\|P_\rmA)+H(Q_i)\Big)\bigg)
&=\sum_{x_i^n\in\calX^n}P_\rmA^n(x_i^n)=1,
\end{align}
and \eqref{concat} follows from the lower bound on the probability of the type class $T_Q^{M(n-1)}$ and the fact that summing over $(M-1)$ concatenated types of length $n$ is equivalent to summing over a type of length $(M-1)n$.

Given any pair of nominal and anomalous distributions $(P_\rmN,P_\rmA)$, we can upper bound the false alarm probability as follows:
\begin{align}
\rmP_{\mathrm{fa}}(\psi_n|P_\rmN,P_\rmA)
&=\bbP_\rmr\{h(\bX^n)>\lambda\}\\
&=\sum_{i\in[M]}\bbP_\rmr\{i^*(X^n)=i\mathrm{~and~}h(\bX^n)>\lambda\}\\
&\leq \sum_{i\in[M]}\bbP_\rmr\{i^*(X^n)=i\mathrm{~and~}\exists~j\in\calM_i:~\rmS_j(X^n)>\lambda\}\label{usedefh}\\
&\leq \sum_{i\in[M]}\sum_{j\in\calM_i}\bbP_\rmr\{\rmS_j(X^n)>\lambda\}\\
&\leq \sum_{i\in[M]}\sum_{j\in\calM_i}\sum_{\bx^n:\rmS_j(\bx^n)>\lambda}\prod_{t\in[M]}P_\rmN(x_t^n)\\
&=\sum_{i\in[M]}\sum_{j\in\calM_i}\sum_{\substack{\bQ\in\calP_n(\calX)^M:\\\rmG_j(\bQ)>\lambda}}\sum_{\substack{\bx^n:~\forall~j\in[M]\\ x_j^n\in\calT_{Q_j}^n}}\exp\left(-n\sum_{t\in[M]}\left( D(Q_t\|P_\rmN)+H(Q_t)\right)\right)\\
&\leq \sum_{i\in[M]}\sum_{j\in\calM_i}\exp(-n\lambda+|\calX|\log((M-1)n+1))\label{simitoerroranalysis}\\
&\leq M(M-1)\exp(-n\lambda+|\calX|\log((M-1)n+1)),
\end{align}
where \eqref{usedefh} follows since when $i^*(X^n)=i$, $h(X^n)=\min_{j\in\calM_i}\rmS_j(X^n)$ and \eqref{simitoerroranalysis} follows from the steps analogously to those leading to the result in \eqref{torefer}.

Finally, we next analyze the false reject probabilities for any $(P_\rmN,P_\rmA)$. For this purpose, we need the following definition of typical sequences for each $i\in[M]$:
\begin{align}
\label{def:typicalset}
\calT_i(P_\rmN,P_\rmA)
&:=\bigg\{\bx^n\in\calX^{Mn}:~\forall~j\in\calM_i,~\|\hatT_{x_j^n}-P_\rmN\|_{\infty}\leq \sqrt{\frac{\log n}{n}}\mathrm{~and~}
\|\hatT_{x_i^n}-P_\rmA\|_{\infty}\leq \sqrt{\frac{\log n}{n}}\bigg\}.
\end{align}
Using Chebyshev's inequality~(c.f.~\cite[Lemma 24]{tan2014state}), we conclude that for each $i\in[M]$,
\begin{align}
\bbP_i\{\bX^n\notin\calT_i(P_\rmN,P_\rmA)\}
&\leq \frac{2M|\calX|}{n^2}=:\mu_n\label{p_atypical}.
\end{align}

In subsequent analysis, we need to use the following properties of $\rmG_i(\bQ)$ (cf. \eqref{def:gi}) for each $i\in[M]$ and any given vector of distributions $\bQ=(Q_1,\ldots,Q_M)\in(\calP(\calX))^M$,
\begin{align}
\frac{\partial \rmG_i(\bQ)}{\partial Q_j(x)}
&=\log\frac{(M-1)Q_j(x)}{\sum_{k\in\calM_i}Q_k(x)},~j\in\calM_i,~x\in\mathrm{supp}(Q_j),\\
\frac{\partial^2 \rmG_i(\bQ)}{\partial (Q_j(x))^2}
&=\frac{\sum_{k\in\calM_i}Q_k(x)-Q_j(x)}{Q_j(x)\Big(\sum_{k\in\calM_i}Q_k(x)\Big)},~j\in\calM_i,~x\in\mathrm{supp}(Q_j),\\
\frac{\partial^2 \rmG_i(\bQ)}{\partial Q_j(x)Q_l(x)}
&=-\frac{1}{\sum_{k\in\calM_i}Q_k(x)},~(j,l)\in\calM_i\times\calM_{i,j}\mathrm{~and~}x\in\mathrm{supp}(Q_j)\cap\mathrm{supp}(Q_l)
\end{align}
For each $i\in[M]$, define the vector of distributions $\bP_i:=(Q_1,\ldots,Q_M)$ with $Q_i=P_\rmA$ and $Q_j=P_\rmN$ for all $j\in\calM_i$. Since KL divergence $D(P_\rmA\|\frac{(M-2)P_\rmN+P_\rmA}{M-1})$ is continuous in its arguments $(P_\rmA,\frac{(M-2)P_\rmN+P_\rmA}{M-1})$ in the interior of simplex, we know that $\rmG_j(\bQ)$ is continuous around $\bP_i$ for each $j\in\calM_i$. Similarly, $\rmG_i(\bQ)$ is continuous around $\bP_i$. Under hypothesis $\rmH_i$, given any set of $M$ sequences $\bx^n\in\calT_i(P_\rmN,P_\rmA)$, one can apply a Taylor expansion of $\rmG_j(\hatT_{x_1^n},\ldots,\hatT_{x_M^n})$ (cf. \eqref{def:scoref}) around $\bP_i$, for each $j\in\calM_i$,  we have
\begin{align}
\nn&\rmG_j(\hatT_{x_1^n},\ldots,\hatT_{x_M^n})\\*
\nn&=D\bigg(P_\rmA\bigg\|\frac{(M-2)P_\rmN+P_\rmA}{M-1}\bigg)+\sum_{x\in\calX}(\hatT_{x_i^n}(x)-P_\rmA(x))\imath_1(x|P_\rmN,P_\rmA)
\\*
\nn&\qquad+\sum_{l\in\calM_{i,j}}\Bigg(
D\bigg(P_\rmN\bigg\|\frac{(M-2)P_\rmN+P_\rmA}{M-1}\bigg)+\sum_{x\in\calX}(\hatT_{x_j^n}(x)-P_\rmN(x))\imath_2(x|P_\rmN,P_\rmA)\Bigg)\\*
&\qquad+\sum_{l\in[M]}O(\|\hatT_{x_j^n}-P_\rmA\|^2)
\\
&=\frac{1}{n}\sum_{t\in[n]}\Big(\imath_1(x_{i,t}|P_\rmN,P_\rmA)+\sum_{l\in\calM_{i,j}}\imath_2(x_{l,t}|P_\rmN,P_\rmA)\Big)+O\left(\frac{\log n}{n}\right)\label{taylor1},
\end{align}
and for $j=i$,
\begin{align}
\rmG_j(\hatT_{x_1^n},\ldots,\hatT_{x_M^n})=O\left(\frac{\log n}{n}\right)\label{taylor2}.
\end{align}

With the above definitions and results, we can now upper bound the false reject probability of our test (cf.~\eqref{test1outlier}) under each hypothesis $\rmH_i$ with $i\in[M]$ with respect to any pair of distributions $(P_\rmN,P_\rmA)$ as follows:
\begin{align}
\nn&\zeta_i(\psi_n|P_\rmN,P_\rmA)\\*
&=\bbP_i\{h(\bX^n)\leq\lambda\}\\
&\leq \bbP_i\Big\{\min_{j\in\calM_i}G_j(\hatT_{X_1^n},\ldots,\hatT_{X_M^n})\leq \lambda\Big\}\label{whyhold}\\
&=1-\bbP_i\Big\{\forall~j\in\calM_i,~G_j(\hatT_{X_1^n},\ldots,\hatT_{X_M^n})>\lambda\Big\}\label{newach1},
\end{align}
where \eqref{whyhold} follows since $h(\bX^n)\geq \min_{j\in\calM_i}G_j(\hatT_{X_1^n},\ldots,\hatT_{X_M^n})$, which is implied by the definition of $h(\bx^n)$ in \eqref{def:defh_smin}.

For simplicity, given random variables $X_1,\ldots,X_M$, for each $i\in[M]$ and $j\in\calM_i$, define the information density
\begin{align}
\imath_{i,j}(X_1,\ldots,X_M|P_\rmN,P_\rmA)
&:=\imath_1(X_i|P_\rmN,P_\rmA)+\sum_{l\in\calM_{i,j}}\imath_2(X_l|P_\rmN,P_\rmA),
\end{align}
and for each $t\in[n]$, we use $\bX_t$ to denote the snapshot of the $M$ sequences at time $t$, i.e., $X_{1,t},\ldots,X_{M,t}$.

The second term in \eqref{newach1} can be lower bounded as follows:
\begin{align}
\nn&\bbP_i\Big\{\forall j\in\calM_i,~G_j(\hatT_{X_1^n},\ldots,\hatT_{X_M^n})>\lambda\Big\}\\*
&\geq \bbP_i\Big\{\forall j\in\calM_i,~G_j(\hatT_{X_1^n},\ldots,\hatT_{X_M^n})>\lambda~\mathrm{~and}~\bX^n\in\calT_i(P_\rmN,P_\rmA)\Big\}\\
&\geq \bbP_i\bigg\{\forall~j\in\calM_i,~\frac{1}{n}\sum_{t\in[n]}\imath_{i,j}(\bX_t|P_\rmN,P_\rmA)>\lambda+O\left(\frac{\log n}{n}\right)~\mathrm{and}~\bX^n\in\calT_i(P_\rmN,P_\rmA)\bigg\}\label{iusemany}\\
&\geq \bbP_i\bigg\{\forall~j\in\calM_i,~\frac{1}{n}\sum_{t\in[n]}\imath_{i,j}(\bX_t|P_\rmN,P_\rmA)>\lambda+O\left(\frac{\log n}{n}\right)\bigg\}-\mu_n\label{newach2},
\end{align}
where \eqref{iusemany} follows from the result in \eqref{p_atypical} and the Taylor expansion in \eqref{taylor1}, and \eqref{newach2} follows from the result in \eqref{p_atypical}.

Recall that under $\bbP_i$, for each $t\in[n]$, $\bX_t=(X_{1,t},\ldots,X_{M_t})$ are independent where $X_{i,t}\sim P_\rmA$ and $X_{j,t}\sim P_\rmN$ for $j\in\calM_i$. Recalling definitions of $\mathrm{GD}_M(P_\rmN,P_\rmA)$in \eqref{def:GD}, $\rmV_M(P_\rmN,P_\rmA)$ in \eqref{def:V} and $\mathrm{Cov}_M(P_\rmN,P_\rmA)$ in \eqref{def:cov}, we have that for any $i\in[M]$ and $j\in\calM_i$
\begin{align}
\mathbb{E}_{\bbP_i}[\imath_{i,j}(\bX_t|P_\rmN,P_\rmA)]
&=\mathrm{GD}_M(P_\rmN,P_\rmA),\label{e1}\\
\mathrm{Var}_{\bbP_i}[\imath_{i,j}(\bX_t|P_\rmN,P_\rmA)]
&=\rmV_M(P_\rmN,P_\rmA)\label{v1},
\end{align}
and for any $k\in\calM_{i,j}$, the covariance of $(\imath_{i,j}(\bX_t|P_\rmN,P_\rmA),\imath_{i,k}(\bX_t|P_\rmN,P_\rmA))$ satisfies
\begin{align}
\mathrm{Cov}_{\bbP_i}[\imath_{i,j}(\bX_t|P_\rmN,P_\rmA)\imath_{i,k}(\bX_t|P_\rmN,P_\rmA)]=\mathrm{Cov}_M(P_\rmN,P_\rmA)\label{calculateCov},
\end{align}
where the justification of \eqref{calculateCov} is provided in Appendix \ref{proof:calulateCov}.

Recall the definition of $\bV_M(P_\rmN,P_\rmA)$ in \eqref{defdispersionmatrix}. Applying the multivariate Berry-Esseen theorem~\cite{Ben03}, the first term in \eqref{newach2} is bounded below as follows:
\begin{align}
\nn&\bbP_i\bigg\{\forall~j\in\calM_i,~\frac{1}{n}\sum_{t\in[n]}\imath_{i,j}(\bX_t|P_\rmN,P_\rmA)>\lambda+O\left(\frac{\log n}{n}\right)\bigg\}\\*
&\geq \rmQ_{M-1}\bigg(\sqrt{n}\Big(\lambda-\mathrm{GD}_M(P_\rmN,P_\rmA)+O\Big(\frac{\log n}{n}\Big)\Big)\times\mathbf{1}_{M-1};\mathbf{0}_{M-1};\bV_M(P_\rmN,P_\rmA)\bigg)+O\left(\frac{1}{\sqrt{n}}\right)\label{newach3},
\end{align}
where $\rmQ_{M-1}(\cdot)$ is the multivariate generalization of the complementary Gaussian cdf defined in \eqref{def:kQ}.

Using \eqref{newach1} and \eqref{newach3}, we have that for any $(P_\rmN,P_\rmA)$, the false reject probability is upper bounded as follows:
\begin{align}
\zeta_i(\psi_n|P_\rmN,P_\rmA)
&\leq 1-\rmQ_{M-1}\bigg(\Big(\lambda-\mathrm{GD}_M(P_\rmN,P_\rmA)+O\Big(\frac{\log n}{n}\Big)\Big)\times\mathbf{1}_{M-1};\mathbf{0}_{M-1};\bV_M(P_\rmN,P_\rmA)\bigg)+O\left(\frac{1}{\sqrt{n}}\right).
\end{align}

\subsection{Proof of Theorem \ref{converse:1outlier}}
\label{proof:converse}
Note that the converse proof without a constraint on the false alarm probability is also a converse proof with a constraint on the false alarm probability. Therefore, in the subsequent proof, we drop the constraint on the false alarm probability and focus on the misclassification error and the false reject probabilities.

We first relate the performances of any test with the type-based test (i.e., a test which uses only the types (empirical distributions) of the sequences $(\hatT_{X_1^n},\ldots,\hatT_{X_M^n})$), as demonstrated in the following lemma.
\begin{lemma}
\label{type:optimal}
Given any test $\phi_n$, for any $\kappa\in[0,1]$, we can construct a type-based test $\phi_n^\rmT$ such that for each $i\in[M]$ and any pair of distributions $(P_\rmN,P_\rmA)$,
\begin{align}
\beta_i(\phi_n|P_\rmN,P_\rmA)&\geq\frac{1-\kappa}{M-1}\beta_i(\phi_n^\rmT|P_\rmN,P_\rmA),\\
\zeta_i(P_\rmN,P_\rmA)&\geq \kappa \zeta_i(\phi_n^\rmT|P_\rmN,P_\rmA).
\end{align}
\end{lemma}
The proof of Lemma \ref{type:optimal} is inspired by \cite[Lemma 2]{gutman1989asymptotically} and \cite[Lemma 5.1]{zhou2018binary} and provided in Appendix \ref{proof:type:optimal}.

We then show that for any type-based test, if we require that the misclassification error probabilities under each hypothesis decay exponentially fast for all pairs of distributions, then the false reject probability under each hypothesis for any particular pair of distributions can be lower bounded by an information spectrum bound, which is the cdf of the second minimal values of the scoring functions $\{\rmG_i(\hatT_{X_1^n},\ldots,\hatT_{X_M^n})\}$.

For simplicity, let
\begin{align}
\eta_{n,M}&:=\frac{M|\calX|\log(n+1)}{n}\label{def:etanm}.
\end{align}
Furthermore, given any tuple of types $\bQ=(Q_1,\ldots,Q_M)\in(\calP_n(\calX))^M$ and any $\lambda\in\bbR_+$, let
\begin{align}
g^*(\bQ)&:=\min_{i\in[M]}\rmG_i(\bQ),\\
g(\bQ)&:=\min_{i\in[M]:\rmG_i(\bQ)>g^*(\bQ)}\rmG_i(\bQ)
\end{align}
denote the minimal and second minimal values of $\{\rmG_i(\bQ)\}_{i\in[M]}$.

\begin{lemma}
\label{type:lb}
Given any $\lambda\in\bbR_+$, for any type-based test $\phi_n^\rmT$ such that for all pair of distributions $(\tilP_\rmN,\tilP_\rmA)$,
\begin{align}
\max_{i\in[M]}\beta_i(\phi_n^\rmT|\tilP_\rmN,\tilP_\rmA)\leq \exp(-n\lambda)\label{err:req},
\end{align}
then for any pair of distributions $(P_\rmN,P_\rmA)$ and for each $i\in[M]$, we have
\begin{align}
\zeta_i(\phi_n^\rmT|P_\rmN,P_\rmA)\geq 
\bbP_i\Big\{g(\hatT_{X_1^n},\ldots,\hatT_{X_M^n})+\eta_{n,M}\le \lambda\Big\}.
\end{align}
\end{lemma}
The proof of Lemma \ref{type:lb} is provided in Appendix \ref{proof:type:lb}.

Combining Lemmas \ref{type:optimal} and \ref{type:lb} with $\kappa=1-\frac{1}{n}$ and noting that $g(\hatT_{x_1^n},\ldots,\hatT_{x_M^n})=h(\bx^n)$ (cf. \eqref{def:defh_smin}) for any $\bx^n=(x_1^n,\ldots,x_M^n)$, we obtain the following corollary.

\begin{corollary}
\label{coro}
Given any $\lambda\in\bbR_+$, for any test $\phi_n$ satisfying that for all pairs of distributions $(\tilP_\rmN,\tilP_\rmA)$
\begin{align}
\max_{i\in[M]}\beta_i(\phi_n|\tilP_\rmN,\tilP_\rmA)\leq \exp(-n\lambda)\label{1out:conrequire},
\end{align}
we have that for any pair of distributions $(P_\rmN,P_\rmA)$ and for each $i\in[M]$
\begin{align}
\zeta_i(\phi_n|P_\rmN,P_\rmA)
&\geq \left(1-\frac{1}{n}\right)\bbP_i\Big\{h(\bX^n)+\eta_{n,M}+\frac{\log n+\log(M-1)}{n}\le \lambda\Big\}.
\end{align}
\end{corollary}

Using Corollary \ref{coro}, with any test $\phi_n$ satisfying \eqref{1out:conrequire},
for any pair of distributions $(P_\rmN,P_\rmA)$, we have that for each $i\in[M]$ and any $j\in\calM_i$, 
\begin{align}
\nn&\zeta_i(\phi_n|P_\rmN,P_\rmA)\\*
\nn&\geq \left(1-\frac{1}{n}\right)\bbP_i\Big\{h(\bX^n)+\eta_{n,M}+\frac{\log n+\log(M-1)}{n}\le \lambda,\\*
&\qquad\qquad\qquad\qquad\qquad\mathrm{and~}h(\bX^n)=\min_{j\in\calM_i}\rmG_{j}(\hatT_{X_1^n},\ldots,\hatT_{X_M^n})\Big\}\\
\nn&\geq \left(1-\frac{1}{n}\right)\bigg(\bbP_i\Big\{\min_{j\in\calM_i}\rmG_{j}(\hatT_{X_1^n},\ldots,\hatT_{X_M^n})+\eta_{n,M}+\frac{\log n+\log(M-1)}{n}\le \lambda\Big\}\\*
&\qquad\qquad\qquad\qquad-\bbP_i\{h(\bX^n)\neq\min_{j\in\calM_i}\rmG_{j}(\hatT_{X_1^n},\ldots,\hatT_{X_M^n})\}
\bigg)\label{newconverse1}.
\end{align}

We first focus on the second term in the bracket of \eqref{newconverse1}. Given any $i\in[M]$, we have that for each $j\in\calM_i$:
\begin{align}
\nn&\bbP_i\{G_j(\hatT_{X_1^n},\ldots,\hatT_{X_M^n})\leq G_i(\hatT_{X_1^n},\ldots,\hatT_{X_M^n})\}\\*
&\leq\bbP_i\{G_j(\hatT_{X_1^n},\ldots,\hatT_{X_M^n})\leq G_i(\hatT_{X_1^n},\ldots,\hatT_{X_M^n}),\bX^n\in\calT_i(P_\rmN,P_\rmA)\}+\bbP_i\{\bX^n\notin\calT_i(P_\rmN,P_\rmA)\}\\
&\leq\bbP\bigg\{\frac{1}{n}\sum_{t\in[n]}\Big(\imath_1(X_{i,t}|P_\rmN,P_\rmA)+\sum_{l\in\calM_{i,j}}\imath_2(X_{l,t}|P_\rmN,P_\rmA)\Big)\leq O\left(\frac{\log n}{n}\right)\bigg\}+\mu_n\label{usetaylorandatypical}\\
&\leq \rmQ\Bigg(\frac{\sqrt{n}(\mathrm{GD}_M(P_\rmN,P_\rmA)+O(\frac{\log n}{n}))}{\sqrt{\rmV_M(P_\rmN,P_\rmA)}}\Bigg)+\frac{6\rmT(P_\rmN,P_\rmA)}{\sqrt{n\big(\rmV_M(P_\rmN,P_\rmA)\big)^3}}+\mu_n\label{usemulberry}\\
&\leq \exp\Bigg(-\frac{n(\mathrm{GD}_M(P_\rmN,P_\rmA)+O(\frac{\log n}{n}))^2}{2\rmV_M(P_\rmN,P_\rmA)}\Bigg)+\frac{6\rmT(P_\rmN,P_\rmA)}{\sqrt{n\big(\rmV_M(P_\rmN,P_\rmA)\big)^3}}+\mu_n\label{useineqrmQ}\\
&=:\kappa_n=O\left(\frac{1}{\sqrt{n}}\right),
\end{align}
where \eqref{usetaylorandatypical} follows from Taylor expansions in \eqref{taylor1} and \eqref{taylor2} and the upper bound on the atypical set in \eqref{p_atypical}, \eqref{usemulberry} follows from the Berry-Esseen theorem~\cite{berry1941accuracy,esseen1942liapounoff} and \eqref{useineqrmQ} follows since $\rmQ(x)\leq \exp(-\frac{x^2}{2})$ for any $x>0$. Therefore, we conclude that for each $i\in[M]$,
\begin{align}
\nn&\bbP_i\{h(\bX^n)\neq\min_{j\in\calM_i}G_j(\hatT_{X_1^n},\ldots,\hatT_{X_M^n})\}\\
&=\bbP_i\Big\{\exists~j\in\calM_i\mathrm{~s.t.~}G_j(\hatT_{X_1^n},\ldots,\hatT_{X_M^n})<G_i(\hatT_{X_1^n},\ldots,\hatT_{X_M^n})\Big\}\\
&\leq\sum_{j\in\calM_i}\bbP_i\Big\{G_j(\hatT_{X_1^n},\ldots,\hatT_{X_M^n})<G_i(\hatT_{x_1^n},\ldots,\hatT_{x_M^n})\Big\}\\
&\leq 1-(M-1)\kappa_n\label{whp_secondmin}.
\end{align}

Finally, analogously to the achievability proof, we analyze the first term in the bracket of \eqref{newconverse1}:
\begin{align}
\nn&\bbP_i\Big\{\min_{j\in\calM_i}\rmG_{j}(\hatT_{X_1^n},\ldots,\hatT_{X_M^n})+\eta_{n,M}+\frac{\log n+\log(M-1)}{n}\le \lambda\Big\}\\*
&=1-\bbP_i\Big\{\forall~j\in\calM_i,~\rmG_{j}(\hatT_{X_1^n},\ldots,\hatT_{X_M^n})+\eta_{n,M}+\frac{\log n+\log(M-1)}{n}>\lambda\Big\}\\
\nn&\geq 1-\bbP_i\Big\{\bX^n\notin\calT_i(P_\rmN,P_\rmA)\}\\
&\qquad-\bbP_i\Big\{\forall~j\in\calM_i,~\rmG_{j}(\hatT_{X_1^n},\ldots,\hatT_{X_M^n})+\eta_{n,M}+\frac{\log n+\log(M-1)}{n}>\lambda\mathrm{~and~}\bX^n\in\calT_i(P_\rmN,P_\rmA)\Big\}\label{usetypical}\\
&\geq 1-\mu_n-\bbP_i\bigg\{\forall~j\in\calM_i~,\frac{1}{n}\sum_{t\in[n]}\imath_{i,j}(\bX_t|P_\rmN,P_\rmA)+O\Big(\frac{\log n}{n}\Big)>\lambda\bigg\}\label{usetaylorcon1out}\\
&\geq 1-\mu_n-\rmQ_{M-1}\bigg(\sqrt{n}\Big(\lambda-\mathrm{GD}_M(P_\rmN,P_\rmA)+O\Big(\frac{\log n}{n}\Big)\Big)\times\mathbf{1}_{M-1};\mathbf{0}_{M-1};\bV_M(P_\rmN,P_\rmA)\bigg)+O\left(\frac{1}{\sqrt{n}}\right)\label{mulBerryagain}
\end{align}
where in \eqref{usetypical}, the definition of the typical set $\calT_i(P_\rmN,P_\rmA)$ was in \eqref{def:typicalset}, \eqref{usetaylorcon1out} follows from the result in \eqref{p_atypical} that upper bounds the probability of $\bbP_i\Big\{\bX^n\notin\calT_i(P_\rmN,P_\rmA)\}$, the Taylor expansion of $\rmG_{j}(\hatT_{X_1^n},\ldots,\hatT_{X_M^n})$ exactly the same as in \eqref{newach2} and the fact that $\eta_{n,M}=O(\log n/n)$, and \eqref{mulBerryagain} follows from the multivariate Berry-Esseen theorem similarly to \eqref{newach3}.

Combining \eqref{newconverse1} and \eqref{mulBerryagain}, we conclude that
\begin{align}
\min_{i\in[M]}\zeta_i(\phi_n|P_\rmN,P_\rmA)
&\geq 1-\rmQ_{M-1}\bigg(\sqrt{n}\Big(\lambda-\mathrm{GD}_M(P_\rmN,P_\rmA)+O\Big(\frac{\log n}{n}\Big)\Big)\times\mathbf{1}_{M-1};\mathbf{0}_{M-1};\bV_M(P_\rmN,P_\rmA)\bigg)+O\left(\frac{1}{\sqrt{n}}\right).
\end{align}
The proof of Theorem \ref{converse:1outlier} is now completed.

\subsection{Proof of Theorem \ref{result:exp1out}}
\label{proof:result:exp1out}

\subsubsection{Achievability}
We make use the same test $\phi_n(\cdot)$ (cf. \eqref{test1outlier} and \eqref{test1:equivalent}) as in the achievability proof of Theorem \ref{result:1outlier}. 

The analyses of the misclassification error probabilities $\beta_i(\psi_n|\tilP_\rmN,\tilP_\rmA)$ and the false alarm probability $\rmP_{\mathrm{fa}}(\phi_n|\tilP_\rmN,\tilP_\rmA)$ are exactly the same as in Appendix \ref{proof:result:1outlier}. It suffices to bound the false reject probability of our test for a particular pair of distributions $(P_\rmN,P_\rmA)$. For each $i\in[M]$, we have that
\begin{align}
\zeta_i(\psi_n|P_\rmN,P_\rmA)
&=\bbP_i\{\phi_n(\bX^n)=\rmH_\rmr\}\\
&=\bbP_i\{h(\bX^n)\leq \lambda\}\\
&=\bbP_i\{\exists~(j,k)\in[M]^2~\mathrm{s.t.~}j\neq k,~\rmS_j(\bX^n)\leq \lambda\mathrm{~and~}\rmS_k(\bX^n)\leq \lambda\}\\
&\leq \sum_{(j,k)\in[M]^2:j\neq k}\bbP_i\{\rmS_j(\bX^n)\leq \lambda\mathrm{~and~}\rmS_k(\bX^n)\leq \lambda\}\\
&\leq \frac{M(M-1)}{2}\max_{(j,k)\in[M]^2:j\neq k}\bbP_i\{\rmS_j(\bX^n)\leq \lambda\mathrm{~and~}\rmS_k(\bX^n)\leq \lambda\}\label{step1:exp}.
\end{align}
We now focus on upper bound the probability term in \eqref{step1:exp}. For any $(j,k)\in[M]^2$, given any $i\in[M]$, we have
\begin{align}
\nn&\bbP_i\{\rmS_j(\bX^n)\leq \lambda\mathrm{~and~}\rmS_k(\bX^n)\leq \lambda\}\\
&\leq\sum_{\bx^n\in\calX^{Mn}:\rmS_j(\bx^n)\leq\lambda,~\rmS_k(\bx^n)\leq\lambda}\bbP_i(\bx^n)\\
&\leq\sum_{\substack{\bQ\in(\calP_n(\calX))^M:\\\rmG_j(\bQ)\leq\lambda,~\rmG_k(\bQ)\leq\lambda}}\exp\bigg(-n\Big(D(Q_i\|P_\rmA)+\sum_{l\in\calM_i}D(Q_j\|P_\rmN)\Big)\bigg)\\
&\leq \sum_{\substack{\bQ\in(\calP_n(\calX))^M}}\exp\bigg(-n\min_{\substack{\bQ\calP_n(\calX))^M:\\\rmG_j(\bQ)\leq\lambda,~\rmG_k(\bQ)\leq\lambda}}\Big(D(Q_i\|P_\rmA)+\sum_{l\in\calM_i}D(Q_j\|P_\rmN)\Big)\bigg)\label{typerej}\\
&\leq (n+1)^{M|\calX|}\exp\bigg(-n\min_{\substack{\bQ\in(\calP(\calX))^M:\\\rmG_j(\bQ)\leq\lambda,~\rmG_k(\bQ)\leq\lambda}}\Big(D(Q_i\|P_\rmA)+\sum_{l\in\calM_i}D(Q_j\|P_\rmN)\Big)\bigg)\label{step2:exp}.
\end{align}
Combining \eqref{step1:exp}, \eqref{step2:exp} and using the definitions of $\mathrm{LD}_i(\cdot)$ in \eqref{def:LDi}, we have that for each $i\in[M]$ and any pair of distributions $(P_\rmN,P_\rmA)$, the $i$-th false reject probability satisfies for any $\lambda\in\bbR_+$
\begin{align}
\liminf_{n\to\infty}-\frac{1}{n}\log\zeta_i(\psi_n|P_\rmN,P_\rmA)
\geq \mathrm{LD}_i(\lambda|P_\rmN,P_\rmA)\label{step3}.
\end{align}

\subsubsection{Converse}
For simplicity, let
\begin{align}
\kappa_{n,M}
&:=\eta_{n,M}+\frac{\log n+\log(M-1)}{n}\label{def:kappanM}.
\end{align}

Using Corollary \ref{coro}, we have that for any test $\phi_n$ such that the misclassification error probabilities decay exponentially fast with speed at least $\lambda$ for all pairs of distributions, given any $(P_\rmN,P_\rmA)$, for each $i\in[M]$, the $i$-th false reject probability $\zeta_i(\phi_n|P_\rmN,P_\rmA)$ satisfies
\begin{align}
\nn&\left(1-\frac{1}{n}\right)\times \zeta_i(\phi_n|P_\rmN,P_\rmA)\\*
&\geq \bbP_i\big\{h(\bX^n)+\kappa_{n,M}\leq \lambda\big\}\\
&=\bbP_i\big\{\exists (j,k)\in[M]^2:~j\neq k,~\rmS_j(\bX^n)+\kappa_{n,M}\leq \lambda\mathrm{~and~}\rmS_k(\bX^n)+\kappa_{n,M}\leq \lambda\big\}\label{def:useh2}\\
&\geq \max_{(j,k)\in[M]^2:~j\neq k}\bbP_i\big\{\rmS_j(\bX^n)+\kappa_{n,M}\leq \lambda\mathrm{~and~}\rmS_k(\bX^n)+\kappa_{n,M}\leq \lambda\big\}\label{choosemax}\\
&\geq (n+1)^{-M|\calX|}\max_{\substack{(j,k)\in[M]^2:\\j\neq k}}\sum_{\substack{\bQ\in(\calP_n(\calX))^M}}\exp\bigg(-n\min_{\substack{\bQ\in(\calP_n(\calX))^M:\\\rmG_j(\bQ)\leq\lambda-\kappa_{n,M}\\\rmG_k(\bQ)\leq\lambda-\kappa_{n,M}}}\Big(D(Q_i\|P_\rmA)+\sum_{l\in\calM_i}D(Q_j\|P_\rmN)\Big)\bigg)\label{exp:constep1},
\end{align}
where \eqref{def:useh2} follows from the definition of $h(\bx^n)$ in \eqref{def:defh_smin} and \eqref{exp:constep1} follows similarly to \eqref{typerej}.

Using the continuity of $(P_\rmN,P_\rmA)$ to $\mathrm{LD}_i(\lambda|P_\rmN,P_\rmA)$ (cf. \eqref{def:LDi}) for any $\lambda\in\bbR_+$, the definition of $\kappa_{n,M}$ in \eqref{def:kappanM} and the results in \eqref{exp:constep1}, we have that for each $i\in[M]$,
\begin{align}
\limsup_{n\to\infty}-\frac{1}{n}\log\zeta_i(\phi_n|P_\rmN,P_\rmA)\leq \mathrm{LD}_i(\lambda|P_\rmN,P_\rmA)
\end{align}
for any test $\phi_n$ satisfying \eqref{concondition}. 

\subsection{Proof of Theorem \ref{result:Toutlier}}
\label{proof:result:Toutlier}

The proof of Theorem \ref{result:Toutlier} is a generalization of the proof of of Theorem \ref{result:1outlier} and thus we only emphasize the differences.

For subsequent analyses, define the following linear combination of third absolute moments
\begin{align}
\rmT(\calB,\calC,P_\rmN,\bP_\calB)
\nn&:=\sum_{i\in(\calB\cap\calM_{\calC})}\mathbb{E}_{P_{\rmA,\jmath_\calB(i)}}[|\imath_{1,\jmath_\calB(i)}(X|\calB,\calC,P_\rmN,\bP_\calB)-\mathbb{E}_{P_{\rmA,\jmath_\calB(i)}}[\imath_{1,\jmath_\calB(i)}(X|\calB,\calC,P_\rmN,\bP_\calB)]|^3]\\*
&\quad+\sum_{i\in(\calM_{\calB}\cap\calM_{\calC})}\mathbb{E}_{P_\rmN}[|\imath_2(X|\calB,\calC,P_\rmN,\bP_\calB)-\mathbb{E}_{P_\rmN}[\imath_2(X|\calB,\calC,P_\rmN,\bP_\calB)]|^3]\label{def:TBC}.
\end{align}
Note that $\rmT(\calB,\calC,P_\rmN,\bP_\calB)$ is finite since we consider distributions $(P_\rmN,\bP_\calB)$ with the same support on the finite alphabet $\calX$.

Recall that $\calS=\bigcup_{t\in[T]}\calS_t$ denotes all possible subsets of $[M]$ with at most $\lceil\frac{M}{2}-1\rceil$ elements. For any $\calC\in\calS$, recall the definition of the scoring function $\rmS_\calC(\bx^n)$ in \eqref{def:scalC}. Recall the definition of the scoring function $\rmS_\calB(\bx^n)=\rmG_\calB(\hatT_{x_1^n},\ldots,\hatT_{x_M^n})$ for any $\calB\in\calS$. Given any $\bx^n$, parallel to \eqref{def:i*_min} and \eqref{def:defh_smin}, define two quantities
\begin{align}
\calI^*(\bx^n)&:=\argmin_{\calB\in\calS}\rmS_\calB(\bx^n)\label{def:setB*},\\
h_\calS(\bx^n)&:=\min_{\calB\in\calS_\calB\neq \calI^*(\bx^n)}\rmS_\calB(\bx^n)\label{def:htbxn}.
\end{align}
Note that $\calI^*(\bx^n)$ denotes the set $\calB$ that minimizes the scoring function and $h_\calS(\bx^n)$ denotes the second minimal value of the scoring function.

The test in \eqref{test:Toutlier} is equivalently expressed as follows:
\begin{align}
\label{test:equaivalent:Toutlier}
\Psi_n(\bx^n)
&=\left\{
\begin{array}{cc}
\rmH_\calB&\mathrm{if~}\calI^*(\bx^n)=\calB,~\mathrm{and~}h_\calS(\bx^n)>\lambda\\
\rmH_\rmr&\mathrm{if~}h_\calS(\bx^n)\leq \lambda.
\end{array}
\right.
\end{align}

We next analyze the performance of the test in \eqref{test:equaivalent:Toutlier}. We first analyze the misclassification error probability. Given any $\calB\in\calS$, under any tuple of distributions $\bP_{\calB}=(P_\rmN,P_{\rmA,1},\ldots,P_{\rmA,|\calB|})$, similarly to the case with at most one outlying sequence, the misclassification error is upper bounded as follows:
\begin{align}
\nn&\beta_{\calB}(\Psi_n|P_\rmN,\bP_T)\\*
&=\bbP_{\calB}\big\{\calI^*(\bX^n)\neq\calB~\mathrm{and~}h_\calS(\bX^n)>\lambda\big\}\\
&\leq \bbP_{\calB}\big\{\rmS_\calB(\bX^n)>\lambda\big\}\\
&=\sum_{\substack{\bQ\in(\calP_n(\calX))^M:\\\rmG_\calB(\bQ)>\lambda}}
\sum_{\substack{\bx^n:x_j^n\in\calT_{Q_j}^n\\\forall j\in[M]}}
\bigg(\prod_{i\in\calB}P_{\rmA,\jmath_\calB(i)}(x_i^n)\bigg)\times\bigg(\prod_{j\in\calM_\calB}P_\rmN(x_j^n)\bigg)\\
&=\sum_{\substack{\bQ\in(\calP_n(\calX))^M:\\\rmG_\calB(\bQ)>\lambda}}
\sum_{\substack{\bx^n:x_j^n\in\calT_{Q_j}^n\\\forall j\in[M]}}\exp\bigg(-n\bigg(\sum_{i\in\calB}D(Q_i\|P_{\rmA,\jmath_\calB(i)})+\sum_{j\in\calM_\calB}D(Q_i\|P_\rmN)+\sum_{i\in[M]}H(Q_i)\bigg)\bigg)\\
\nn&=\sum_{\substack{\bQ\in(\calP_n(\calX))^M:\\\rmG_\calB(\bQ)>\lambda}}
\sum_{\substack{\bx^n:x_j^n\in\calT_{Q_j}^n\\\forall j\in[M]}}\exp\bigg(-n\bigg(\sum_{i\in\calB}D(Q_i\|P_{\rmA,\jmath_\calB(i)})+\sum_{i\in[M]}H(Q_i)\\
&\qquad\qquad\qquad\qquad\qquad\qquad\qquad+\rmG_\calB(\bQ)+(M-|\calB|)D\left(\frac{\sum_{k\in\calM_\calB}Q_k}{M-T}\bigg\|P_\rmN\right)\bigg)\bigg)\\
&\leq \exp(-n\lambda)\sum_{Q_j\in\calP_n(\calX),~j\in\calM_\calB}\exp\bigg(-n(M-|\calB|)D\left(\frac{\sum_{k\in\calM_\calB}Q_k}{M-T}\bigg\|P_\rmN\right)\bigg)\\
&\leq\exp\Big(-n\lambda+|\calX|\log((M-|\calB|)n+1)\Big)\label{mtofollow}\\
&\leq \exp\Big(-n\lambda+|\calX|\log((M-1)n+1)\Big),
\end{align}
We then analyze the false alarm probability. Given any nominal distribution $P_\rmN$, the false alarm probability is upper bounded as follows:
\begin{align}
\rmP_{\mathrm{fa}}(\Psi_n|P_\rmN,\bP_T)
&:=\bbP_\rmr\{h_\calS(\bX^n)>\lambda\}\\
&=\sum_{\calB\in\calS}\bbP_\rmr\{\calI^*(\bX^n)=\calB\mathrm{~and~}h_\calS(\bX^n)>\lambda\}\\
&\leq \sum_{\calB\in\calS}\bbP_\rmr\{\calI^*(\bX^n)=\calB\mathrm{~and~}\exists~\calC\in\calS:~\calC\neq\calB,~\rmS_{\calC}(\bX^n)>\lambda\}\\
&\leq \sum_{\calB\in\calS}\sum_{\calC\in\calS:~\calC\neq\calB}\bbP_\rmr\{\rmS_{\calC}(\bX^n)>\lambda\}\\
&\leq \sum_{\calB\in\calS}\sum_{\calC\in\calS:~\calC\neq\calB}\exp\Big(-n\lambda+|\calX|\log((M-1)n+1)\Big)\label{mfollow2}\\
&\leq |\calS|^2\exp\Big(-n\lambda+|\calX|\log((M-1)n+1)\Big)
\end{align}
where \eqref{mfollow2} follows from similar steps leading to \eqref{mtofollow}.

Finally, we analyze the false reject probability of the tests. For this purpose, we need a generalized version of the typical set in \eqref{def:typicalset}. For each $\calB\in\calS$ and any $\bP_\calB$,  define
\begin{align}
\label{typical:Toutlier}
\calT_\calB(\bP_\calB)
\nn&:=\bigg\{\bx^n\in\calX^{Mn}:\forall~j\in\calB,~\|\hatT_{x_j^n}-P_{\rmA,\jmath_\calB(j)}\|_{\infty}\leq\sqrt{\frac{\log n}{n}},\\*
&\qquad\qquad\mathrm{~and}~\forall j\in\calM_\calB,\|\hatT_{x_j^n}-P_\rmN\|_{\infty}\leq\sqrt{\frac{\log n}{n}}\bigg\}.
\end{align}

Similarly to \eqref{p_atypical}, for each $\calB\in\calS$, we have
\begin{align}
\bbP_\calB\{\bX^n\notin\calT_\calB(\bP_\calB)\}\leq \frac{2M|\calX|}{n^2}\label{p_atypical2}.
\end{align}

Recall the definitions of the mixture distribution in \eqref{def:Pmix} and the information densities in \eqref{def:i1T} and \eqref{def:i2T}. Under each hypothesis $H_\calB$, given any observed sequences $\bx^n\in\calT_\calB(\bP_\calB)$, applying Taylor expansions of $\rmG_\calC(\hatT_{x_1^n},\ldots,\hatT_{x_M^n})$ for $\calC\in\calS$ around $\bP_\calB$ yields
\begin{itemize}
\item if $\calC\neq \calB$, then
\begin{align}
\nn&\rmG_\calC(\hatT_{x_1^n},\ldots,\hatT_{x_M^n})\\*
\nn&=\sum_{j\in(\calB\cap\calM_\calC)}\Big(D(P_{\rmA,\jmath_\calB(j)}\|P_{\rm{Mix}}^{\calB,\calC,P_\rmN,\bP_\calB})+\sum_{x}\big(\hatT_{x_j^n}(x)-P_{\rmA,\jmath_\calB(j)}(x)\big)\imath_{1,\jmath_\calB(j)}(x|\calB,\calC,P_\rmN,\bP_\calB)+O\left(\|\hatT_{x_j^n}-P_{\rmA,\jmath_\calB(j)}\|^2\right)\Big)\\
&\qquad+\sum_{j\in(\calM_\calB\cap\calM_\calC)}\Big(D(P_\rmN\|P_{\rm{Mix}}^{\calB,\calC,P_\rmN,\bP_\calB})+\sum_{x}\big(\hatT_{x_j^n}(x)-P_\rmN(x)\big)\imath_2(x|\calB,\calB,\bP_\calB)+O\left(\|\hatT_{x_j^n}-P_\rmN\|^2\right)\Big)\\
&=\frac{1}{n}\sum_{t\in[n]}\Big(\sum_{j\in(\calB\cap\calM_\calC)}\imath_{1,\jmath_\calB(j)}(x_{j,t}|\calB,\calC,P_\rmN,\bP_\calB)+\sum_{j\in(\calM_\calB\cap\calM_\calC)}\imath_2(x_{j,t})\Big)+O\left(\frac{\log n}{n}\right)\label{taylorcase1};
\end{align}
\item if $\calC=\calB$, then
\begin{align}
\rmG_\calC(\hatT_{x_1^n},\ldots,\hatT_{x_M^n})=O\left(\frac{\log n}{n}\right)\label{taylorcase2}.
\end{align}
\end{itemize}

The false reject probability is then upper bounded as follows:
\begin{align}
\nn&\zeta_\calB(\Psi_n|P_\rmN,\bP_T)\\*
&=\bbP_\calB\{h_\calS(\bX^n)\leq \lambda\}\\
&\leq \bbP_\calB\Big\{\min_{\calC\in\calS_\calB}\rmG_\calC(\hatT_{X_1^n},\ldots,\hatT_{X_M^n})\leq \lambda\Big\}\\
&=1- \bbP_\calB\Big\{\forall~\calC\in\calS_\calB:\rmG_\calC(\hatT_{X_1^n},\ldots,\hatT_{X_M^n})>\lambda\Big\}\label{Tout:step3},
\end{align}
where $\calS_\calB=\{\calC\in\calS_\calB\}$ denotes the set of sets in $\calS$ that are not equal to $\calB$. We now analyze the probability term in \eqref{Tout:step3}. Recall that given any $(\calB,\calC)\in\calS^2$ and any variables $(X_1,\ldots,X_M)$,
\begin{align}
\imath_{\calB,\calC}(X_1,\ldots,X_M|P_\rmN,\bP_\calB)
&=\sum_{j\in(\calB\cap\calM_\calC)}\imath_{1,\jmath_\calB(j)}(X_j|\calB,\calC,P_\rmN,\bP_\calB)+\sum_{\barj\in(\calM_\calB\cap\calM_\calC)}\imath_2(X_{\barj}|\calB,\calC|P_\rmN,\bP_\calB).
\end{align}
For each $t\in[n]$, we use $\bX_t$ to denote $X_{1,t},\ldots,X_{M,t}$.

Using Taylor expansions in \eqref{taylorcase1}, \eqref{taylorcase2}, for any $\calB\in\calS$,
\begin{align}
\nn&\bbP_\calB\{\forall~\calC\in\calS_\calB:\rmG_\calC(\hatT_{X_1^n},\ldots,\hatT_{X_M^n})>\lambda\}\\*
&\geq \bbP_\calB\{\forall~\calC\in\calS_\calB:\rmG_\calC(\hatT_{X_1^n},\ldots,\hatT_{X_M^n})> \lambda,\bX^n\in\calT_\calB(\bP_\calB)\}\\
&\geq \bbP_\calB\bigg\{\forall~\calC\in\calS_\calB:\frac{1}{n}\sum_{t\in[n]}\imath_{\calB,\calC}(\bX_t|P_\rmN,\bP_\calB)>\lambda+O\left(\frac{\log n}{n}\right)\bigg\}-\bbP_\calB\{\bX^n\notin\calT_\calB(\bP_\calB)\}\\
&=\bbP_\calB\bigg\{\forall~\calC\in\calS_\calB:\frac{1}{n}\sum_{t\in[n]}\imath_{\calB,\calC}(\bX_t|P_\rmN,\bP_\calB)>\lambda+O\left(\frac{\log n}{n}\right)\bigg\}-\frac{2M|\calX|}{n^2}\label{Tout:step4}.
\end{align}

Note that $\calS_\calB$ denotes all subsets of $[M]$ with size no greater than $T$ excluding the set $\calB$ thus each element in $\calS_\calB$ is a subset of $[M]$. There are in total $|\calS|-1$ elements in the set $\calS_\calB$ and thus $|\calS|-1$ inequalities \eqref{Tout:step4} that need to be satisfied simultaneously. Recall that the elements in $\calS_\calB$ are ordered as $\{\calC_1,\ldots,\calC_{|\calS|-1}\}$. This way, the probability term in \eqref{Tout:step4} is equivalent to 
\begin{align}
\bbP_\calB\bigg\{\forall~i\in[|\calS|-1]:\frac{1}{n}\sum_{t\in[n]}\imath_{\calB,\calC_i}(\bX_t|P_\rmN,\bP_\calB)>\lambda+O\left(\frac{\log n}{n}\right)\bigg\}\label{tout:equi}.
\end{align}

Recall the definitions of $\mathrm{GD}(\calB,\calC,P_\rmN,\bP_\calB)$ in \eqref{def:GDBC}, $\rmV(\calB,\calC,P_\rmN,\bP_\calB)$ in \eqref{def:VBC}. Note that for each $i\in[|\calS|-1]$ and each $t\in[n]$,
\begin{align}
\mathbb{E}_{\bbP_\calB}[\imath_{\calB,\calC_i}(\bX_t|P_\rmN,\bP_\calB)]
&=\mathrm{GD}(\calB,\calC_i,P_\rmN,\bP_\calB),\\
\mathrm{Var}_{\bbP_\calB}[\imath_{\calB,\calC_i}(\bX_t|P_\rmN,\bP_\calB)]
&=\rmV(\calB,\calC_i,P_\rmN,\bP_\calB).
\end{align}
Furthermore, for any $(i,k)\in[|\calS|-1]^2$ such that $i\neq k$, we have
\begin{align}
\mathrm{Cov}_{\bbP_\calB}(\imath_{\calB,\calC_i}(\bX_t|P_{\calB}),\imath_{\calB,\calD_j}(\bX_t|P_{\calB}))=\mathrm{Cov}(\calC_i,\calC_j,\bP_\calB)\label{toverify}.
\end{align}

Using \eqref{Tout:step4} to \eqref{toverify}, and applying the multivariate Berry-Esseen theorem similarly to \eqref{newach3}, we have
\begin{align}
\nn&\bbP_\calB\{\forall~\calC\in\calS_\calB:\rmG_\calC(\hatT_{X_1^n},\ldots,\hatT_{X_M^n})> \lambda\}\\*
&\geq \rmQ_{|\calS|-1}\big(\sqrt{n}\bar{\mu}(\lambda,P_\rmN,\bP_\calB);\mathbf{0}_{|\calS|-1};\bV(\calB,P_\rmN,\bP_\calB)\big)+O\left(\frac{1}{\sqrt{n}}\right)\label{mberryresult},
\end{align}
where $\bar{\mu}(\lambda,P_\rmN,\bP_\calB)$ denotes the vector $(\lambda-\mathrm{GD}(\calB,\calC_1,P_\rmN,\bP_\calB)+O(\log n/n),\ldots,\lambda-\mathrm{GD}(\calB,\calC_{|\calS|-1},P_\rmN,\bP_\calB)+O(\log n/n))$.

Combining \eqref{Tout:step3} and \eqref{mberryresult}, we conclude that for any $\calB\in\calS$, the false reject probability under hypothesis $\rmH_\calB$ satisfies that
\begin{align}
\zeta_\calB(\Psi_n|P_\rmN,\bP_T)\leq 1-\rmQ_{|\calS|-1}\big(\sqrt{n}\bar{\mu}(\lambda,P_\rmN,\bP_\calB);\mathbf{0}_{|\calS|-1};\bV(\calB,P_\rmN,\bP_\calB)\big)+O\left(\frac{1}{\sqrt{n}}\right).
\end{align}
The proof of Theorem \ref{result:Toutlier} is completed.

\subsection{Proof of Theorem \ref{converse:m}}
\label{converse:toutlier}
Recall the definition of $h_\rmT(\bx^n)$ in \eqref{def:htbxn} and $\eta_{n,M}$ in \eqref{def:etanm}. For ease of notation, let
\begin{align}
\eta_{n,M,T}&:=\eta_{n,M}+\frac{\log n+\log(|\calS|))}{n}.
\end{align}
The following corollary is key to the converse proof of Theorem \ref{result:Toutlier}.
\begin{corollary}
\label{coro:Tout}
Given any $\lambda\in\bbR_+$, for any test $\phi_n$ such that for all tuples of nominal and anomalous distributions $(\tilP_\rmN,\tilde{\bP}_T)$,
\begin{align}
\beta_\calB(\phi_n|\tilP_\rmN,\tilde{\bP}_T)\leq \exp(-n\lambda)\label{Tout:conreqire},
\end{align}
then for any tuple of nominal and anomalous distributions $(P_\rmN,\bP_T)$, for each $\calB\in\calS$, 
\begin{align}
\zeta_\calB(\phi_n|P_\rmN,\bP_T)\geq
\left(1-\frac{1}{n}\right)\bbP_\calB\big\{h_\calS(\bX^n)+\eta_{n,M,T}\leq\lambda\big\}.
\end{align}
\end{corollary}
The proof of Corollary \ref{coro:Tout} is similar to that of Corollary \ref{coro} and is thus omitted.

Using Corollary \ref{coro:Tout}, for any test $\phi_n$ satisfying \eqref{Tout:conreqire}, given any tuple of distributions $(P_\rmN,\bP_T)$, for each $\calB\in\calS$, the false reject probability is lower bounded by
\begin{align}
\nn&\zeta_\calB(\phi_n|P_\rmN,\bP_T)\\*
&\geq \left(1-\frac{1}{n}\right)\bbP_\calB\big\{h_\calS(\bX^n)+\eta_{n,M,T}\leq\lambda,~h_\calS(\bX^n)=\min_{\calC\in\calS_\calB}\rmG_{\calC}(\hatT_{X_1^n},\ldots,\hatT_{X_M^n})\big\}\\
\nn&\geq \left(1-\frac{1}{n}\right)\bbP_\calB\big\{\min_{\calC\in\calS_\calB}\rmG_{\calC}(\hatT_{X_1^n},\ldots,\hatT_{X_M^n})+\eta_{n,M,T}\leq\lambda\big\}\\*
&\qquad\qquad-\bbP_\calB\big\{h_\calS(\bX^n)\neq\min_{\calC\in\calS_\calB}\rmG_{\calC}(\hatT_{X_1^n},\ldots,\hatT_{X_M^n})\big\}\\
&\geq \left(1-\frac{1}{n}\right)\bbP_\calB\big\{\min_{\calC\in\calS_\calB}\rmG_{\calC}(\hatT_{X_1^n},\ldots,\hatT_{X_M^n})+\eta_{n,M,T}\leq\lambda\big\}+O\left(\frac{1}{\sqrt{n}}\right)\label{usestep2},
\end{align}
where \eqref{usestep2} is justified in Appendix \ref{proof:2:0}.

Analogous to \eqref{mberryresult}, using the multivariate Berry-Esseen theorem, we conclude that 
\begin{align}
\nn&\bbP_\calB\big\{\min_{\calC\in\calS_\calB}\rmG_{\calC}(\hatT_{X_1^n},\ldots,\hatT_{X_M^n})+\eta_{n,M,T}\leq\lambda\big\}\\*
&=1-\bbP_\calB\big\{\forall~\calC\in\calS_\calB,~\rmG_{\calC}(\hatT_{X_1^n},\ldots,\hatT_{X_M^n})+\eta_{n,M,T}>\lambda\big\}\\
&\geq 1-\rmQ_{|\calS|-1}\big(\sqrt{n}\bar{\mu}(\lambda,P_\rmN,\bP_\calB);\mathbf{0}_{|\calS|-1};\bV(\calB,P_\rmN,\bP_\calB)\big)+O\left(\frac{1}{\sqrt{n}}\right)\label{mconverse:last}.
\end{align}
The proof of Theorem \ref{converse:m} is thus completed by combining \eqref{usestep2} and \eqref{mconverse:last}.

\subsection{Proof of Lemma \ref{type:optimal}}
\label{proof:type:optimal}

For simplicity, let $\bQ:=(Q_1,\ldots,Q_M)\in(\calP_n(\calX))^M$ and for any $\bQ$, we use $\calT_{\bQ}^n$ to denote the set of sequences $\bx=(x_1^n,\ldots,x_M^n)$ such that $x_i^n\in\calT_{Q_i}^n$ for all $i\in[M]$. Given any test $\phi_n$, the sample space $\calX^{Mn}$ is separated into $(M+1)$ disjoint regions: $\{\calA_i(\phi_n)\}_{i\in[M]}$ and $\calA_\rmr(\phi_n)$ where
\begin{align}
\calA_i(\phi_n)&=\{\bx\in\calX^{Mn}:\phi_n(\bx)=\rmH_i\},\\
\calA_\rmr(\phi_n)&=\Big(\bigcup_{i\in[M]}\calA_i\Big)^\rmc.
\end{align}

We can then construct a type-based test as follows. Given any $\kappa$, for any $\bQ\in(\calP_n(\calX))^M$, 
\begin{itemize}
\item $\phi_n^\rmT(\bQ)=\rmH_\rmr$ if at least $\kappa$ fractions of the sequences in the type class $\calT_{\bQ}^n$ are contained in the reject region, i.e.,
\begin{align}
|\calT_{\bQ}^n\cap\calA_\rmr(\phi_n)|\geq \kappa|\calT_{\bQ}^n|.
\end{align}
\item $\phi_n^\rmT(\bQ)=\rmH_i$ if i) less than $\kappa$ fractions of the sequences in the type class $\calT_{\bQ}^n$ are contained in the reject region and ii) for all $j\in[M]$, $\calA_i(\phi_n)$ contains the most number of the sequences in the type class $\calT_{\bQ}^n$, i.e.,
\begin{align}
|\calT_{\bQ}^n\cap\calA_\rmr(\phi_n)|<\kappa|\calT_{\bQ}^n|,~\mathrm{and~}|\calT_{\bQ}^n\cap\calA_i(\phi_n)|\geq \max_{j\in\calM_i}|\calT_{\bQ}^n\cap\calA_j(\phi_n)|.
\end{align}
\end{itemize}

For any pair of distributions $(P_\rmN,P_\rmA)$, we can then relate the performances of an arbitrary test $\phi_n$ and the constructed type-based test $\phi_n^\rmT$ as follows:
\begin{align}
\beta_i(\phi_n|P_\rmT,P_\rmA)
&=\bbP_i\Big\{\bigcup_{j\in\calM_i}\calA_j(\phi_n)\Big\}\\
&=\sum_{j\in\calM_i}\bbP\{\calA_j(\phi_n)\}\\
&=\sum_{j\in\calM_i}\sum_{\bQ\in(\calP_n(\calX))^M}\bbP_i\{\calA_j(\phi_n)\cap\calT_{\bQ}^n\}\\
&\geq \sum_{j\in\calM_i}\sum_{\substack{\bQ\in(\calP_n(\calX))^M:|\calT_{\bQ}^n\cap\calA_\rmr(\phi_n)|<\kappa\\|\calT_{\bQ}^n\cap\calA_j(\phi_n)|\geq\max_{l\in\calM_j}
|\calT_{\bQ}^n\cap\calA_j(\phi_n)|}}\bbP_i\{\calA_j(\phi_n)\cap\calT_{\bQ}^n\}\\
&\geq \sum_{j\in\calM_i}\sum_{\substack{\bQ\in(\calP_n(\calX))^M:\phi_n^\rmT(\bQ)=\rmH_j}}\frac{1-\kappa}{M-1}\bbP_i\{\calT_{\bQ}^n\}\\
&=\frac{1-\kappa}{M-1}\sum_{\substack{\bQ\in(\calP_n(\calX))^M:\exists j\in\calM_i:\phi_n^\rmT(\bQ)=\rmH_j}}\bbP_i\{\calT_{\bQ}^n\}\\
&=\frac{1-\kappa}{M-1}\beta_i(\phi_n^\rmT|P_1,P_2),
\end{align}
and
\begin{align}
\zeta_i(\phi_n|P_\rmT,P_\rmA)
&=\bbP_i\Big\{\calA_\rmr(\phi_n)\Big\}\\
&=\sum_{\bQ\in(\calP_n(\calX))^M}\bbP_i\{\calA_\rmr(\phi_n)\cap\calT_{\bQ}^n\}\\
&\geq \sum_{\substack{\bQ\in(\calP_n(\calX))^M:|\calT_{\bQ}^n\cap\calA_\rmr(\phi_n)|\geq\kappa}}\bbP_i\{\calA_j(\phi_n)\cap\calT_{\bQ}^n\}\\
&\geq \kappa \sum_{\substack{\bQ\in(\calP_n(\calX))^M:|\calT_{\bQ}^n\cap\calA_\rmr(\phi_n)|\geq\kappa}}\bbP_i\{\calT_{\bQ}^n\}\\
&=\kappa\zeta_i(\phi_n^\rmT|P_\rmT,P_\rmA).
\end{align}

\subsection{Proof of Lemma \ref{type:lb}}
\label{proof:type:lb}

To prove Lemma \ref{type:lb}, it suffices to prove that for any type-based test satisfying \eqref{err:req}, if a tuple of types $\bQ$ satisfies that
\begin{align}
g(\bQ)+\eta_{n,M}<\lambda\label{rej:con},
\end{align}
then we must have $\phi_n^\rmT(\bQ)=\rmH_\rmr$.

We will prove our claim by contradiction. Suppose our claim is not true. Then there exist types $\bar{\bQ}=(\barQ_1,\ldots,\barQ_M)\in(\calP_n(\calX))^M$ such that for some $i\in[M]$,
\begin{align}
\phi_n^\rmT(\bar{\bQ})=\rmH_i~\mathrm{and~}g(\bar{\bQ})+\eta_{n,M}<\lambda\label{toprovecon}
\end{align}

Note that \eqref{toprovecon} implies that there exists $(j,k)\in\calM^2$ such that $j\neq k$ and 
\begin{align}
\rmG_j(\bar{\bQ})+\eta_{n,M}<\lambda~\mathrm{and~}\rmG_k(\bar{\bQ})+\eta_{n,M}<\lambda.
\end{align}
Furthermore, either $j\neq i$ or $k\neq i$. Without loss of generality, we assume that $j\neq i$.

Then, we have that for all pairs of distributions $(\tilP_\rmN,\tilP_\rmA)$, the misclassification error probability under hypothesis $\rmH_j$ can be lower bounded as follows:
\begin{align}
\beta_j(\phi_n^\rmT|\tilP_\rmN,\tilP_\rmA)
&\geq \sum_{\bQ\in(\calP_n(\calX))^M:\phi_n^\rmT(\bQ)=\rmH_i}\bbP_j(\calT_{\bQ}^n)\\
&\geq \bbP_j(T_{\bar{\bQ}}^n)\\
&\geq (n+1)^{-Mn}\exp\Big(-n\big(D(\barQ_j\|\tilP_\rmA)+\sum_{l\in\calM_j}D(\barQ_l\|\tilP_\rmN)\big)\Big).
\end{align}
Now if we let $\tilP_\rmA=\barQ_j$ and $\tilP_\rmN=\frac{\sum_{l\in\calM_i}\barQ_l}{M-1}$, then
\begin{align}
\beta_j(\phi_n^\rmT|\tilP_\rmN,\tilP_\rmA)
&\geq (n+1)^{-M}\exp(-n\rmG_i(\bar{\bQ}))\\
&=\exp\Big(-n(\rmG_i(\bar{\bQ})+\eta_{n,M})\Big)\\
&>\exp(-n\lambda),
\end{align}
which contradicts that \eqref{err:req} holds. Therefore, we have show that for any type-based test $\phi_n^\rmT$ satisfying \eqref{err:req}, we must have $\phi_n^\rmT(\bQ)=\rmH_\rmr$ for any $\bQ$ satisfying \eqref{rej:con}.

\subsection{Justification of Properties of Exponent Tradeoff}
\label{sec:just}

We first prove that $\mathrm{LD}_i(\lambda,P_\rmN,P_\rmA)=0$ if and only if $\lambda\geq \mathrm{GD}_M(P_\rmN,P_\rmA)$. Recall the definition of $\mathrm{LD}_i(\cdot)$ in \eqref{def:LDi} and the definition of $\rmG_i(\cdot)$ in \eqref{def:gi}. Note that for each $i\in[M]$, any $\lambda\in\bbR_+$ and any $(P_\rmN,P_\rmA)$, $\mathrm{LD}_i(\lambda,P_\rmN,P_\rmA)=0$ if there exists $(j,k)\in[M]^2$ such that $j\neq k$, $\rmG_j(\bQ^*)\leq \lambda$ and $\rmG_k(\bQ^*)\leq \lambda$ where $\bQ^*$ is a collection of distributions with $Q_i^*=P_\rmA$ and $Q_t^*=P_\rmN$ for all $t\in\calM_i$. For any $j\in\calM_i$, we have
\begin{align}
\rmG_j(\bQ^*)
&=\sum_{t\in(\calM_i\cap\calM_j)}D\left(Q_t^*\bigg\|\frac{\sum_{t\in\calM_j}Q_l}{M-1}\right)+D\left(Q_i^*\bigg\|\frac{\sum_{t\in\calM_j}Q_l}{M-1}\right)\\
&=(M-2)D\left(P_\rmN\bigg\|\frac{(M-2)P_\rmN+P_\rmA}{M-1}\right)+D\left(P_\rmA\bigg\|\frac{(M-2)P_\rmN+P_\rmA}{M-1}\right)\\
&=\mathrm{GD}_M(P_\rmN,P_\rmA)\label{just1}.
\end{align}
Furthermore, if $j=i$, then 
\begin{align}
\rmG_j(\bQ^*)
&=\sum_{t\in \calM_i}D\left(Q_t^*\bigg\|\frac{\sum_{t\in\calM_j}Q_l}{M-1}\right)=0\label{just2}.
\end{align}

Combining \eqref{just1} and \eqref{just2}, we have for each $i\in[M]$,
\begin{align}
\max_{(j,k)\in[M]^2:j\neq k}\max\{\rmG_j(\bQ^*),\rmG_k(\bQ^*)\}=\mathrm{GD}_M(P_\rmN,P_\rmA).
\end{align}
Therefore, if $\lambda=\mathrm{GD}_M(P_\rmN,P_\rmA)$, for each $i\in[M]$, we can find $(j,k)\in[M]^2$ such that $j\neq k$, $\max\{\rmG_j(\bQ^*),\rmG_k(\bQ^*)\}\leq \lambda$ and thus $\mathrm{LD}_i(\lambda,P_\rmN,P_\rmA)=0$. The justification is completed by the above argument with the fact that $\mathrm{LD}_i(\lambda,P_\rmN,P_\rmA)$ is non-increasing in $\lambda$ for each $i\in[M]$ and any $(P_\rmN,P_\rmA)$.

We then prove \eqref{toprove2}. Since $\mathrm{LD}_i(\lambda,P_\rmN,P_\rmA)$ is non-increasing in $\lambda$ for each $i\in[M]$ and any $(P_\rmN,P_\rmA)$, then we have
\begin{align}
\nn&\sup_{\lambda\in\bbR_+}\mathrm{LD}_i(\lambda,P_\rmN,P_\rmA)\\
&\leq \mathrm{LD}_i(0,P_\rmN,P_\rmA)\\
&=\min_{(j,k)\in[M]^2:j\neq k}\min_{\substack{\bQ\in(\calP(\calX))^M:\\\rmG_j(\bQ)=0,~\rmG_k(\bQ)=0}} \Big(D(Q_i\|P_\rmA)+\sum_{l\in\calM_i}D(Q_l\|P_\rmN)\Big)\\
&=\min_{(j,k)\in[M]^2:j\neq k}\min_{\substack{\bQ\in(\calP(\calX))^M:\\Q_1=Q_2=\ldots Q_M}} \Big(D(Q_i\|P_\rmA)+\sum_{l\in\calM_i}D(Q_l\|P_\rmN)\Big)\label{useGi}\\
&=\min_{Q\in\calP(\calX)} \Big(D(Q\|P_\rmA)+(M-1)D(Q\|P_\rmN)\Big),
\end{align}
where \eqref{useGi} follows from the definition of $\rmG_i(\cdot)$ in \eqref{def:gi}. The proof of \eqref{toprove2} is thus completed.

\subsection{Justification of \eqref{calculateCov}}
\label{proof:calulateCov}

For any $i\in[M]$, $j\in\calM_i$, $k\in\calM_{i,j}$, given any pair of distributions $(P_\rmN,P_\rmA)$,
\begin{align}
\nn&\mathrm{Cov}_{\bbP_i}[\imath_{i,j}(\bX_t|P_\rmN,P_\rmA)\imath_{i,k}(\bX_t|P_\rmN,P_\rmA)]\\*
&=\mathbb{E}_{\bbP_i}[\imath_{i,j}(\bX_t|P_\rmN,P_\rmA)\imath_{i,k}(\bX_t|P_\rmN,P_\rmA)]-\mathbb{E}_{\bbP_i}[\imath_{i,j}(\bX_t|P_\rmN,P_\rmA)]\mathbb{E}_{\bbP_i}[\imath_{i,k}(\bX_t|P_\rmN,P_\rmA)]\\
&=\mathbb{E}_{\bbP_i}[\imath_{i,j}(\bX_t|P_\rmN,P_\rmA)\imath_{i,k}(\bX_t|P_\rmN,P_\rmA)]-\big(\mathrm{GD}_M(P_\rmN,P_\rmA)\big)^2,\label{covcontinue}
\end{align}
where \eqref{covcontinue} follows from \eqref{e1}. The first term in \eqref{covcontinue} can be further calculated as follows:
\begin{align}
\nn&\mathbb{E}_{\bbP_i}[\imath_{i,j}(\bX_t|P_\rmN,P_\rmA)\imath_{i,k}(\bX_t|P_\rmN,P_\rmA)]\\*
&=\mathbb{E}_{\bbP_i}\Big[\big(\imath_1(X_{i,t}|P_\rmN,P_\rmA)+\sum_{l\in\calM_{i,j}}\imath_2(X_{l,t}|P_\rmN,P_\rmA)\big)\big(\imath_1(X_{i,t}|P_\rmN,P_\rmA)+\sum_{\barl\in\calM_{i,k}}\imath_2(X_{\barl,t}|P_\rmN,P_\rmA)\big)\Big]\\
\nn&=\mathbb{E}_{\bbP_i}\Big[\big(\imath_1(X_{i,t}|P_\rmN,P_\rmA)\big)^2\Big]+\mathbb{E}_{\bbP_i}\Big[\sum_{\barl\in\calM_{i,k}}\imath_1(X_{i,t}|P_\rmN,P_\rmA)\imath_2(X_{\barl,t}|P_\rmN,P_\rmA)\Big]\\
\nn&\qquad+\mathbb{E}_{\bbP_i}\Big[\sum_{l\in\calM_{i,j}}\imath_1(X_{i,t}|P_\rmN,P_\rmA)\imath_2(X_{l,t}|P_\rmN,P_\rmA)\Big]\\
&\qquad+\mathbb{E}_{\bbP_i}\Big[\big(\sum_{l\in\calM_{i,j}}\imath_2(X_{l,t}|P_\rmN,P_\rmA)\big)\big(\sum_{\barl\in\calM_{i,k}}\imath_2(X_{\barl,t}|P_\rmN,P_\rmA)\big)\Big]\label{covcontinue2}.
\end{align}
We can calculate each term in \eqref{covcontinue2}. The first term in \eqref{covcontinue2} satisfies
\begin{align}
\mathbb{E}_{\bbP_i}\Big[\big(\imath_1(X_{i,t}|P_\rmN,P_\rmA)\big)^2\Big]
&=\mathbb{E}_{P_\rmA}\Big[\big(\imath_1(X|P_\rmN,P_\rmA)\big)^2\Big]\label{term1}.
\end{align}

The second term in \eqref{covcontinue2} satisfies
\begin{align}
\mathbb{E}_{\bbP_i}\Big[\sum_{\barl\in\calM_{i,k}}\imath_1(X_{i,t}|P_\rmN,P_\rmA)\imath_2(X_{\barl,t}|P_\rmN,P_\rmA)\Big]
&=\sum_{\barl\in\calM_{i,k}}\mathbb{E}_{\bbP_i}[\imath_1(X_{i,t}|P_\rmN,P_\rmA)]\mathbb{E}_{\bbP_i}[\imath_2(X_{\barl,t}|P_\rmN,P_\rmA)]\\
&=(M-2)\mathbb{E}_{P_\rmA}[\imath_1(X|P_\rmN,P_\rmA)]\mathbb{E}_{P_\rmN}[\imath_2(X|P_\rmN,P_\rmA)]
\label{term2}.
\end{align}
Similarly, the third term in \eqref{covcontinue2} satisfies
\begin{align}
\mathbb{E}_{\bbP_i}\Big[\sum_{l\in\calM_{i,j}}\imath_1(X_{k,t}|P_\rmN,P_\rmA)\imath_2(X_{l,t}|P_\rmN,P_\rmA)\Big]
&=(M-2)\mathbb{E}_{P_\rmA}[\imath_1(X|P_\rmN,P_\rmA)]\mathbb{E}_{P_\rmN}[\imath_2(X|P_\rmN,P_\rmA)]\label{term3}.
\end{align}
Finally, the last term in \eqref{covcontinue2} satisfies
\begin{align}
\nn&\mathbb{E}_{\bbP_i}\Big[\big(\sum_{l\in\calM_{i,j}}\imath_2(X_{l,t}|P_\rmN,P_\rmA)\big)\big(\sum_{\barl\in\calM_{i,k}}\imath_2(X_{\barl,t}|P_\rmN,P_\rmA)\big)\Big]\\*
&=\sum_{l\in\calM_{i,j}}\mathbb{E}_{\bbP_i}\Big[\imath_2(X_{l,t}|P_\rmN,P_\rmA)\big(\sum_{\barl\in\calM_{i,k}}\imath_2(X_{\barl,t}|P_\rmN,P_\rmA)\big)\Big]\\
\nn&=\mathbb{E}_{\bbP_i}\Big[\imath_2(X_{k,t}|P_\rmN,P_\rmA)\big(\sum_{\barl\in\calM_{i,k}}\imath_2(X_{\barl,t}|P_\rmN,P_\rmA)\big)\Big]\\*
&\qquad+\sum_{l\in\calM_{i,j,k}}\mathbb{E}_{\bbP_i}\Big[\imath_2(X_{l,t}|P_\rmN,P_\rmA)\big(\sum_{\barl\in\calM_{i,k}}\imath_2(X_{\barl,t}|P_\rmN,P_\rmA)\big)\Big]\\
\nn&=\sum_{\barl\in\calM_{i,k}}\mathbb{E}_{\bbP_i}[\imath_2(X_{k,t}|P_\rmN,P_\rmA)]\mathbb{E}_{\bbP_i}[\imath_2(X_{\barl,t}|P_\rmN,P_\rmA)]\\*
&\qquad+\sum_{l\in\calM_{i,j,k}}\mathbb{E}_{\bbP_i}\Big[\big(\imath_2(X_{l,t}|P_\rmN,P_\rmA)\big)^2+\imath_2(X_{l,t}|P_\rmN,P_\rmA)\big(\sum_{\barl\in\calM_{i,k,l}}\imath_2(X_{\barl,t}|P_\rmN,P_\rmA)\big)\Big]\\
\nn&=(M-2)\big(\mathbb{E}_{P_\rmN}[\imath_2(X|P_\rmN,P_\rmA)]\big)^2+(M-3)\mathbb{E}_{P_\rmN}\Big[\big(\imath_2(X|P_\rmN,P_\rmA)\big)^2\Big]\\*
&\qquad+(M-3)^2\big(\mathbb{E}_{P_\rmN}[\imath_2(X|P_\rmN,P_\rmA)]\big)^2\\
&=(M^2-5M+7)\big(\mathbb{E}_{P_\rmN}[\imath_2(X|P_\rmN,P_\rmA)]\big)^2+(M-3)\mathbb{E}_{P_\rmN}\Big[\big(\imath_2(X|P_\rmN,P_\rmA)\big)^2\Big]
\label{term4}.
\end{align}

Combining \eqref{covcontinue} to \eqref{term4}, we have that for any $i\in[M]$, $j\in\calM_i$, $k\in\calM_{i,j}$,
\begin{align}
\nn&\mathrm{Cov}_{\bbP_i}[\imath_{i,j}(\bX_t|P_\rmN,P_\rmA)\imath_{i,k}(\bX_t|P_\rmN,P_\rmA)]\\
\nn&=-\big(\mathrm{GD}_M(P_\rmN,P_\rmA)\big)^2+\mathbb{E}_{P_\rmA}\Big[\big(\imath_1(X|P_\rmN,P_\rmA)\big)^2\Big]+2(M-2)\mathbb{E}_{P_\rmA}[\imath_1(X|P_\rmN,P_\rmA)]\mathbb{E}_{P_\rmN}[\imath_2(X|P_\rmN,P_\rmA)]\\
&\qquad+(M^2-5M+7)\big(\mathbb{E}_{P_\rmN}[\imath_2(X|P_\rmN,P_\rmA)]\big)^2+(M-3)\mathbb{E}_{P_\rmN}\Big[\big(\imath_2(X|P_\rmN,P_\rmA)\big)^2\Big].
\end{align}

\subsection{Justification of \eqref{usestep2}}
\label{proof:2:0}
Given any $\calB\in\calS$, using the Berry-Esseen theorem and Taylor expansions in \eqref{taylorcase1}, \eqref{taylorcase2}, we have that for each $\calC\in\calS_\calB$
\begin{align}
\nn&\bbP_\calB\{\rmG_\calC(\hatT_{X_1^n},\ldots,\hatT_{X_M^n})<\rmG_\calB(\hatT_{X_1^n},\ldots,\hatT_{X_M^n})\}\\
&\leq \bbP_\calB\{\rmG_\calC(\hatT_{X_1^n},\ldots,\hatT_{X_M^n})<\rmG_\calB(\hatT_{X_1^n},\ldots,\hatT_{X_M^n}),\bX^n\in\calT_\calB(\bP_\calB)\}+\bbP_\calB\{\bX^n\notin\calT_\calB(\bP_\calB)\}\\
&\leq \bbP_\calB\bigg\{\frac{1}{n}\sum_{t\in[n]}\Big(\sum_{j\in(\calB\cap\calM_\calC)}\imath_{1,\jmath_\calB(j)}(X_{j,t}|\calB,\calC,P_\rmN,\bP_\calB)+\sum_{j\in(\calM_\calB\cap\calM_\calC)}\imath_2(X_{j,t})\Big)<O\left(\frac{\log n}{n}\right)\bigg\}+\frac{2M|\calX|}{n^2}\label{useatypical}\\
&\leq \rmQ\Bigg(\frac{\sqrt{n}(\mathrm{GD}(\calB,\calC,P_\rmN,\bP_\calB)+O(\frac{\log n}{n}))}{\sqrt{\rmV(\calB,\calC,P_\rmN,\bP_\calB)}}\Bigg)+\frac{6\rmT(\calB,\calC,P_\rmN,\bP_\calB)}{\sqrt{n(\rmV(\calB,\calC,P_\rmN,\bP_\calB))^3}}+\frac{2M|\calX|}{n^2}\\
&\leq \exp\Bigg(-\frac{n(\mathrm{GD}(\calB,\calC,P_\rmN,\bP_\calB)+O(\frac{\log n}{n}))^2}{2\rmV(\calB,\calC,P_\rmN,\bP_\calB)}\Bigg)+\frac{6\rmT(\calB,\calC,P_\rmN,\bP_\calB)}{\sqrt{n(\rmV(\calB,\calC,P_\rmN,\bP_\calB))^3}}+\frac{2M|\calX|}{n^2}\\
&=:\kappa_{T,n}=O\left(\frac{1}{\sqrt{n}}\right)\label{Tout:step1},
\end{align} 
where \eqref{useatypical} follows from \eqref{p_atypical2}.

Using \eqref{Tout:step1}, we have that for any $\calB\in\calS$,
\begin{align}
\nn&\bbP_\calB\{h_\calS(\bX^n)=\min_{\calC\in\calS_\calB}\rmG_\calC(\hatT_{X_1^n},\ldots,\hatT_{X_M^n})\}\\
&=\bbP_\calB\{\rmG_\calB(\hatT_{X_1^n},\ldots,\hatT_{X_M^n})\leq \min_{\calC\in\calS_\calB}\rmG_\calC(\hatT_{X_1^n},\ldots,\hatT_{X_M^n})\}\\
&\geq 1-\sum_{\calC\in\calS_\calB}\bbP_\calB\{\rmG_\calC(\hatT_{X_1^n},\ldots,\hatT_{X_M^n})<\rmG_\calB(\hatT_{X_1^n},\ldots,\hatT_{X_M^n})\}\\
&\geq 1-(|\calS|-1)\kappa_{T,n}\label{Tout:step2:0}\\
&=1-O\left(\frac{1}{\sqrt{n}}\right)\label{Tout:step2}.
\end{align}

\subsection*{Acknowledgement}
The authors acknowledge anonymous reviewers for many helpful comments and useful suggestions that help improve the quality of the current paper significantly.

\bibliographystyle{IEEEtran}
\bibliography{IEEEfull_lin}
\end{document}

%% file: preamble.tex
\usepackage[mathscr]{eucal}
\usepackage[cmex10]{amsmath}
\usepackage{epsfig,epsf,psfrag}
\usepackage{amssymb,amsmath,amsthm,amsfonts,latexsym}
\usepackage{amsmath,graphicx,bm,xcolor,url,overpic}
\usepackage{fixltx2e}
\usepackage{array}
\usepackage{verbatim}
\usepackage{bm}
\usepackage{algorithmic}
\usepackage{algorithm}
\usepackage{verbatim}
\usepackage{textcomp}
\usepackage{mathrsfs}
\usepackage{epstopdf}

\newcommand{\openone}{\leavevmode\hbox{\small1\normalsize\kern-.33em1}}

\catcode`~=11 \def\UrlSpecials{\do\~{\kern -.15em\lower .7ex\hbox{~}\kern .04em}} \catcode`~=13 

\allowdisplaybreaks[4]

\newcommand{\nn}{\nonumber}

\newcommand{\calA}{\mathcal{A}}
\newcommand{\calB}{\mathcal{B}}
\newcommand{\calC}{\mathcal{C}}
\newcommand{\calD}{\mathcal{D}}

\newcommand{\calI}{\mathcal{I}}

\newcommand{\calM}{\mathcal{M}}
\newcommand{\calN}{\mathcal{N}}

\newcommand{\calP}{\mathcal{P}}

\newcommand{\calS}{\mathcal{S}}
\newcommand{\calT}{\mathcal{T}}

\newcommand{\calX}{\mathcal{X}}

\newcommand{\bP}{\mathbf{P}}

\newcommand{\bQ}{\mathbf{Q}}

\newcommand{\bV}{\mathbf{V}}

\newcommand{\bx}{\mathbf{x}}
\newcommand{\bX}{\mathbf{X}}

\newcommand{\rmA}{\mathrm{A}}

\newcommand{\rmc}{\mathrm{c}}

\newcommand{\rmd}{\mathrm{d}}

\newcommand{\rmG}{\mathrm{G}}

\newcommand{\rmH}{\mathrm{H}}

\newcommand{\rmN}{\mathrm{N}}

\newcommand{\rmP}{\mathrm{P}}

\newcommand{\rmQ}{\mathrm{Q}}
\newcommand{\rmr}{\mathrm{r}}

\newcommand{\rmS}{\mathrm{S}}

\newcommand{\rmT}{\mathrm{T}}

\newcommand{\rmV}{\mathrm{V}}

\newcommand{\rmx}{\mathrm{x}}

\newcommand{\bbE}{\mathsf{E}}

\newcommand{\bbN}{\mathbb{N}}

\newcommand{\bbP}{\mathbb{P}}

\newcommand{\bbR}{\mathbb{R}}

\DeclareMathAlphabet{\mathbsf}{OT1}{cmss}{bx}{n}
\DeclareMathAlphabet{\mathssf}{OT1}{cmss}{m}{sl}

\DeclareSymbolFont{bsfletters}{OT1}{cmss}{bx}{n}  
\DeclareSymbolFont{ssfletters}{OT1}{cmss}{m}{n}
\DeclareMathSymbol{\bsfGamma}{0}{bsfletters}{'000}
\DeclareMathSymbol{\ssfGamma}{0}{ssfletters}{'000}
\DeclareMathSymbol{\bsfDelta}{0}{bsfletters}{'001}
\DeclareMathSymbol{\ssfDelta}{0}{ssfletters}{'001}
\DeclareMathSymbol{\bsfTheta}{0}{bsfletters}{'002}
\DeclareMathSymbol{\ssfTheta}{0}{ssfletters}{'002}
\DeclareMathSymbol{\bsfLambda}{0}{bsfletters}{'003}
\DeclareMathSymbol{\ssfLambda}{0}{ssfletters}{'003}
\DeclareMathSymbol{\bsfXi}{0}{bsfletters}{'004}
\DeclareMathSymbol{\ssfXi}{0}{ssfletters}{'004}
\DeclareMathSymbol{\bsfPi}{0}{bsfletters}{'005}
\DeclareMathSymbol{\ssfPi}{0}{ssfletters}{'005}
\DeclareMathSymbol{\bsfSigma}{0}{bsfletters}{'006}
\DeclareMathSymbol{\ssfSigma}{0}{ssfletters}{'006}
\DeclareMathSymbol{\bsfUpsilon}{0}{bsfletters}{'007}
\DeclareMathSymbol{\ssfUpsilon}{0}{ssfletters}{'007}
\DeclareMathSymbol{\bsfPhi}{0}{bsfletters}{'010}
\DeclareMathSymbol{\ssfPhi}{0}{ssfletters}{'010}
\DeclareMathSymbol{\bsfPsi}{0}{bsfletters}{'011}
\DeclareMathSymbol{\ssfPsi}{0}{ssfletters}{'011}
\DeclareMathSymbol{\bsfOmega}{0}{bsfletters}{'012}
\DeclareMathSymbol{\ssfOmega}{0}{ssfletters}{'012}

\newcommand{\tilP}{\tilde{P}}

\newcommand{\hatT}{\hat{T}}

\newcommand{\barj}{\bar{j}}

\newcommand{\barl}{\bar{l}}

\newcommand{\barQ}{\bar{Q}}

\newcommand{\bmu}{\bm{\mu}}

\newcommand{\bSigma	}{\bm{\Sigma}}

\DeclareMathOperator*{\argmin}{arg\,min}

\newtheorem{theorem}{Theorem} 
\newtheorem{lemma}{Lemma}

\newtheorem{corollary}{Corollary}